\documentclass[12pt]{article}
\usepackage{amssymb}
\def\lsim{\mathrel{\rlap {\raise.5ex\hbox{$ < $}}
{\lower.5ex\hbox{$\sim$}}}}
\def\gsim{\mathrel{\rlap {\raise.5ex\hbox{$ > $}}
{\lower.5ex\hbox{$\sim$}}}}
\def\sqr#1#2{{\vcenter{\vbox{\hrule height.#2pt
        \hbox{\vrule width.#2pt height#1pt \kern#1pt
           \vrule width.#2pt}
        \hrule height.#2pt}}}}

\def\lsim{{\displaystyle
{{\raise-8pt\hbox{$ <$}}
\atop{\raise5pt\hbox{$\sim$}}}}}
\def\gsim{{\displaystyle
{{\raise-8pt\hbox{$ >$}}
\atop{\raise5pt\hbox{$\sim$}}}}}
\def\slsim{{\displaystyle
{{\raise-8pt\hbox{$\scriptstyle <$}}
\atop{\raise5pt\hbox{$\scriptstyle \sim$}}}}}
\def\sgsim{{\displaystyle
{{\raise-8pt\hbox{$\scriptstyle  >$}}
\atop{\raise5pt\hbox{$\scriptstyle \sim$}}}}}

\newcommand{\sump}[0]{\sum_{(h,g)}\!{\raise 4pt \hbox{$'$}}\,}
\newcommand{\Sump}[0]{\sum_{(H,G)}\! \!   {\raise
4pt  \hbox{$'$}}\,}
\newcommand{\sumpf}[0]{\sum_{(H^{\rm f},G^{\rm f})}\! \! \! \! 
{\raise 4pt \hbox{$'$}}\,}
\catcode`\@=11
\newcount\hour
\newcount\minute
\newtoks\amorpm
\hour=\time\divide\hour by60
\minute=\time{\multiply\hour by60 \global\advance\minute by-\hour}
\edef\standardtime{{\ifnum\hour<12 \global\amorpm={am}%
        \else\global\amorpm={pm}\advance\hour by-12 \fi
        \ifnum\hour=0 \hour=12 \fi
        \number\hour:\ifnum\minute<10 0\fi\number\minute\the\amorpm}}
\edef\militarytime{\number\hour:\ifnum\minute<10 0\fi\number\minute}
\def\draftlabel#1{{\@bsphack\if@filesw {\let\thepage\relax
   \xdef\@gtempa{\write\@auxout{\string
      \newlabel{#1}{{\@currentlabel}{\thepage}}}}}\@gtempa
   \if@nobreak \ifvmode\nobreak\fi\fi\fi\@esphack}
        \gdef\@eqnlabel{#1}}
\def\@eqnlabel{}
\def\@vacuum{}
\def\draftmarginnote#1{\marginpar{\raggedright\scriptsize\tt#1}}
\def\draft{\oddsidemargin -.2truein
        \def\@oddfoot{\sl preliminary draft \hfil
        \rm\thepage\hfil\sl\today\quad\militarytime}
        \let\@evenfoot\@oddfoot \overfullrule 3pt
        \let\label=\draftlabel
        \let\marginnote=\draftmarginnote
   \def\@eqnnum{(\theequation)\rlap{\kern\marginparsep\tt\@eqnlabel}%
\global\let\@eqnlabel\@vacuum}  }
\relax
\def\Im{\,{\rm Im}\, }
\def\Re{\,{\rm Re}\, }

\newcommand{\ar}[2]{{#1 \atopwithdelims[]#2}}
\def\a{\alpha}

\def\thefootnote{\fnsymbol{footnote}}
\def\be{\begin{equation}}
\def\ee{\end{equation}}
\def\ba{\begin{eqnarray}}
\def\ea{\end{eqnarray}}
\def\bs{\begin{subequations}}
\def\es{\end{subequations}}

\def\th{\vartheta}

\def\l{\lambda}

\def\o{\omega}
\def\c{\chi}
\def\p{\partial}

\def\t{\tau}

\def\t{\tau}

\def\b{\beta}
\def\ifd{\int_{\cal F}\frac{d^2\tau}{\im}}

\def\ee{\end{equation}}
\def\ba{\begin{eqnarray}}
\def\ea{\end{eqnarray}}
\newtoks\@stequation
\def\limit#1#2{\smash { \mathop{#1} \limits_{#2} }  }
\def\subequations{\refstepcounter{equation}%
  \edef\@savedequation{\the\c@equation}%
  \@stequation=\expandafter{\theequation}
  \edef\@savedtheequation{\the\@stequation}
  \edef\oldtheequation{\theequation}%
  \setcounter{equation}{0}%
  \def\theequation{\oldtheequation\alph{equation}}}
\def\endsubequations{\setcounter{equation}{\@savedequation}%
  \@stequation=\expandafter{\@savedtheequation}%
  \edef\theequation{\the\@stequation}\global\@ignoretrue
  \vspace*{-12pt} \\}
\def\bs{\begin{subequations}}
\def\es{\end{subequations}}
\def\nn{\nonumber}

\def\ifd{\int_{\cal F}\frac{d^2\tau}{\Im\tau}}

\def\np#1#2#3{Nucl. Phys. {\bf{B#1}} (#2) #3}
\def\pl#1#2#3{Phys. Lett. {\bf{B#1}} (#2) #3}

\def\thebibliography#1{%
\vskip 0.5cm \centerline{\bf References}
\list{%
[\arabic{enumi}]}{\settowidth\labelwidth{[#1]}
\leftmargin\labelwidth
\advance\leftmargin\labelsep
\usecounter{enumi}}
\def\newblock{\hskip .11em plus .33em minus .07em}
\sloppy\clubpenalty4000\widowpenalty4000
\sfcode`\.=1000\relax}

\begin{document}
\renewcommand{\theequation}{\arabic{section}.\arabic{equation}}
\begin{titlepage}
\begin{flushright}
CERN-TH/99-08 \\ NEIP-99-001 \\ IOA-99-01 \\
LPTENS/99-01\\
hep-th/9901123 \\
\end{flushright}
\begin{centering}
\vspace{.3in}
{\bf \large Classification of the $N=2$, $Z_2 \times Z_2$-symmetric type II
orbifolds and their type II asymmetric duals}\\
\vspace{1. cm}
{A. GREGORI$^{\ 1}$, C. KOUNNAS$^{\ 2,\, \ast}$ and
J. RIZOS$^{\ 3}$} \\
\medskip
\vskip 1cm
{\it $^1 $ Institut de Physique Th\'{e}orique, Universit\'{e}
de Neuch\^{a}tel \\
2000 Neuch{\^a}tel, Switzerland}\\
\medskip
{\it $^2 $ Theory Division, CERN}\\
{\it 1211 Geneva 23, Switzerland}\\
\medskip
{\it and}\\
\medskip
{\it $^3 $ Division of Theoretical Physics, Physics Department,
University of Ioannina}\\
{\it 45110 Ioannina, Greece}\\
\vspace{1.0cm}
{\bf Abstract}\\
\end{centering}
\vspace{.1in}
Using free world-sheet fermions, we construct and classify all the $N=2$, 
$Z_2 \times Z_2$ four-dimensional orbifolds of the type IIA/B strings 
for which the orbifold projections act symmetrically on the left and right 
movers. We study the deformations of these models out of the fermionic point,
deriving the partition functions at a generic point in the moduli of the 
internal torus $T^6=T^2 \times T^2 \times T^2$. We investigate some of their 
perturbative and non-perturbative dualities and construct new dual pairs
of type IIA/type II asymmetric orbifolds, which are
related non-perturbatively and allow us to gain
insight into some of the non-perturbative properties of the type IIA/B
strings in four dimensions.
In particular, we consider some of the (non-)perturbative
gravitational corrections.
\vspace{.8cm}
\begin{flushleft}
CERN-TH/99-08 \\
January 1999
\end{flushleft}
\hrule width 6.7cm \vskip.1mm{\small \small \small
$^\ast$\ On leave from {\it Laboratoire de Physique Th{\'e}orique de
l'Ecole Normale Sup{\'e}rieure, CNRS}, 
24 rue Lhomond, 75231 Paris Cedex 05, France.}
\end{titlepage}
\newpage
\setcounter{footnote}{0}
\renewcommand{\thefootnote}{\arabic{footnote}}

\setcounter{section}{1}
\section*{\normalsize{\bf 1. Introduction}}

During the recent years, duality has played a fundamental role in 
the progress of string theory.
However, despite the huge amount of work done in this field,
in most of the cases duality has remained a conjecture,
based more on field theory and supersymmetry/supergravity considerations
than on tests made directly at the string level \cite{vw}.
Actually, in order to perform string-loop computations,
it is necessary to go to special points of the moduli space, in which
it is possible to solve the two-dimensional conformal field theory.
In this paper, we study a class of four-dimensional compactifications
of type II strings, with two space-time supersymmetries, for which 
this is possible.
Our main interest is in what we call $Z_2 \times Z_2$ symmetric
orbifolds, namely orbifold constructions in which the $N=8$
supersymmetry is reduced to $N=2$ by two $Z_2$ projections that act
symmetrically on the left and right movers. 
These orbifolds are of particular interest because they can be easily 
realized through a free fermion construction \cite{klt}--\cite{fk}.
In this framework, the various constraints and requirements of a
consistent string theory construction are collected in a set of rules,
which can be easily handled. In particular, we show that it is possible
to write a general formula for the GSO projections, which allows us to 
give a complete classification of such orbifolds.

All of these constructions can be seen as
compactifications on singular limits of CY manifolds.
For all the models, the scalar manifolds are coset spaces:
\be
{SU(1,1)\over U(1)}\times {SO(2,2+N_V)\over SO(2)\times SO(2+N_V)}
\ \ {\rm and}
\ \ {SO(4,4+N_H)\over SO(4)\times SO(4+N_H)},
\ee
describing respectively the space of the $N_V+3$ moduli in the vector
multiplets and that of the $N_H+4$ in the hypermultiplets.
For each pair $(N_V,N_H)$ there always exists a construction 
for which $N_V$ and $N_H$ are exchanged, corresponding
to a compactification on the mirror manifold.
Some of them, namely the models with $N_V=N_H=16,8,0$,
correspond to compactifications on CY manifolds already investigated,
although in slightly different contexts \cite{gm}--\cite{sv}.

For each model, we write 
the (one-loop) partition function, 
which encodes all the information about its perturbative physics.  
We then establish the exact equivalence, for this class of orbifolds,
of the fermionic construction and a geometric 
construction based on bosons compactified at special radii.
In this way, we show that,
once these orbifolds are constructed at the fermionic point,
it is possible to switch on some moduli, namely the moduli $T^i$, $U^i$,
$i=1,2,3$, which on  type IIA are 
associated respectively with the K\"{a}hler class moduli 
and the complex structure moduli
of the three tori into which the compact space
is factorized by the orbifold projections.
The derivation of the partition functions, for any such construction,
at a generic point in the space of these moduli, constitutes
one of the main results of this paper.
This allows us to investigate some deformations of the models.
In particular, from the analysis of certain helicity supertraces
\cite{bk}--\cite{n=6}, which distinguish between various BPS 
and non-BPS states, we read off the presence of perturbative
Higgs and super-Higgs phenomena. The first
account for the appearance of new 
massless states in particular corners of the moduli space, 
while the second, besides that,
determine the restoration of a
certain number of supersymmetries.
Such properties play a key role in the search for dual constructions.
In particular, it is possible to recognize which models correspond to 
compactifications on orbifold limits of K3 fibrations \cite{k3,al}.
In these cases, the heterotic dual constructions can be easily 
identified \cite{fhsv}--\cite{gkp2}.
In this paper, we focus our attention on the duality between  
type IIA/B and type II ``asymmetric'' constructions, in
which all the supersymmetries come from the left movers only, 
the supersymmetries of the other 
chirality being projected out by $(-)^{\rm F_{\rm R}}$, the right 
fermion number operator.
In these constructions, as in the heterotic $N=2$ compactifications,
the dilaton--axion field belongs to a vector multiplet, and
the type IIA/type II asymmetric dual pairs are related by a $U$-duality
similar to the duality of the 
type IIA/heterotic strings 
(examples of such dual pairs were previously
considered in \cite{sv,gkp2}). 
This implies that a perturbative computation performed on one side
gives information on the non-perturbative physics of the dual.
In this paper we present in detail the construction of such 
type II asymmetric duals. Then, as in \cite{gkp,gkp2},
we consider $R^2$ corrections, which serve both as a test of duality
and as the actual computation of a quantity that is non-perturbative
in the type II asymmetric duals.
On the other hand, an investigation of the perturbative super-Higgs
phenomena present in the type II asymmetric models 
tells us about the presence of an 
analogous phenomenon also in the type IIA/B duals, in which
it is entirely
non-perturbative  and could not be seen from an analysis of the 
helicity supertraces.

The paper is organized as follows:

In Section 2 we present the fermionic construction of the $N=2$,
type IIA/B  $Z_2 \times Z_2$ symmetric orbifolds and we discuss
the analysis of the massless spectrum. The reader can find
a short reminder of the rules of the fermionic construction in Appendix A,
while more details on the massless spectrum are given in Appendix B.
At the end of the section we explain our method of classification
of such constructions, quoting in Appendix C the general formulae for the GSO
projections.

In Section 3 we derive the partition functions of the various models.
We establish the equivalence of world-sheet fermions and bosons,
thereby deriving the partition functions at a generic point in the toroidal 
moduli $T^i$, $U^i$. The classification of the partition functions is given
in Appendix D.

In Section 4 we compute the helicity supertraces
and interpret the various orbifold operations in terms of
stringy Higgs and super-Higgs phenomena.

In Section 5 we discuss the mirror symmetry, in the context of these 
symmetric orbifolds, and the non-perturbative dualities relating 
some of these models to heterotic duals 
and/or to type II asymmetric constructions.
The type IIA/type II asymmetric dual pairs are discussed in detail in
Sections 5.2 and  5.3, where we quote also the partition function 
for the type II asymmetric orbifolds.
We discuss the corrections to the $R^2$ term.
The dual pairs are then compared in
Section 5.4, in which we discuss some of their non-perturbative aspects.
A detailed discussion of shifted lattice sums and their integrals over 
the fundamental domain is in Appendix E, while in Appendix F
we discuss the computation of helicity supertraces for the 
type II asymmetric orbifolds.

Our comments and conclusions are given in Section 6.

\noindent

\vskip 0.3cm
\setcounter{section}{2}
\setcounter{equation}{0}
\section*{\normalsize{\bf 2. Type II $N=2$ symmetric orbifolds 
in the fermionic construction}}

The use of free world-sheet fermions
turns out to be convenient for the analysis 
of the massless spectrum and the general classification
of the $Z_2 \times Z_2$ type II
symmetric orbifolds.
In order to construct them, we start from
the $N=8$ type II string,
which is described, in the light-cone gauge, by 8 world-sheet left/right moving
bosonic and fermionic coordinates, $X^{L,R}_i$ and $\psi^{L,R}_i$
$(i=1,...,8)$. In our notation, the coordinates $\psi^{L,R}_{\mu}$
and $X^{L,R}_{\mu}$  $(\mu=1,2)$ represent the space-time 
transverse degrees of freedom, whereas the remaining ones correspond
to the internal degrees of freedom. The $N=8$ string has therefore four 
space-time supersymmetries originating from the left-moving sector and
four from the right-moving sector.
We then introduce $Z_2$ projections, which act symmetrically on 
left/right moving coordinates, reducing the number of supersymmetries
to $N=2$, one coming from the left and one from the right movers.
In the fermionic construction \cite{klt}--\cite{ab}, the
$X^{L,R}_i$, $(i=3,...,8)$ are replaced by
the pairs of Majorana--Weyl spinors $\o_I^{L,R}$ 
and $y_I^{L,R}$, $(I=1,...,6)$.
To follow the standard notation of the fermionic construction \cite{fk},
we rename the internal components of the fields  $\psi^{L,R}_i$
as $\c_I^{L,R}$. The construction of string models then amounts to
a choice of boundary conditions for the fermions, which satisfies local and 
global consistency requirements. A model is defined by a basis of sets 
$\alpha_i$ $(i=1,...,n)$ of fermions and by a modular-invariant choice of 
${n(n-1)/2+1}$ phases (modular coefficients) 
$C_{(\alpha_i|\alpha_j)}$, which determine the GSO 
projections (we refer the reader to  Appendix A for more details).
In this language, the $N=8$ model is constructed by introducing
three basis sets, 
namely $F$, which contains all the left- and right-moving fermions:
\be 
F=\left\{ \begin{array}{l} \psi_{\mu}^L, \c_I^L, y_I^L, \o_I^L \\
        \psi_{\mu}^R, \c_I^R, y_I^R, \o_I^R \end{array}
              \right\}, 
~~~(\mu=1,2;~I=1,...,6),
\ee
and the sets $S$ and $\bar{S}$, which contain only eight left- or
right-moving fermions, and distinguish the boundary
conditions of the left- and right- moving world-sheet superpartners:
\be
S=\left\{ \psi_{\mu}^L, \c_1^L, \ldots,\c_6^L \right\},~~~~~  
\bar{S}=\left\{ \psi_{\mu}^R, \c_1^R,\ldots,\c_6^R \right\}.
\ee
In order to obtain a $Z_2 \times Z_2$ symmetric orbifold, we add to the 
basis the two sets $b_1$ and $b_2$:
\ba
b_{1} & = & 
      \left\{ \begin{array}{l} \psi_{\mu}^L,\c_{1,2}^L,y_{3,\ldots,6}^L \\
       \psi_{\mu}^R,\c_{1,2}^R,y_{3,\ldots,6}^R \end{array}
              \right\}~, 
\label{b1} \\
&& \nn \\
b_{2}& = & 
  \left\{ \begin{array}{l} \psi_{\mu}^L,\c_{3,4}^L,y_{1,2}^L,y_{5,6}^L \\
       \psi_{\mu}^R,\c_{3,4}^R,y_{1,2}^R,y_{5,6}^R \end{array}
              \right\}~. 
\label{b2}
\ea
These sets assign $Z_2$ boundary conditions, thereby introducing
new projections, which break the $N=8$ supersymmetry.

The definition of the model is completed by the choice of
the following modular 
coefficients, which fix the GSO projections and determine the chirality
of the spinors: 
\[
\begin{tabular} {| l |r|r|r|r|r|} \hline 
 & $F$ & $S$ & $\bar{S}$ & $b_1$ & $b_2$ \rule[-.2cm]{0cm}{.7cm} \\ \hline
$F$    & 1 & $-1$ & $-1$ & 1 & 1 \rule[-.2cm]{0cm}{.7cm} \\ \hline
$S$    & $-1$ & 1 &  1 & $-1$ & $-1$ \rule[-.2cm]{0cm}{.7cm} \\ \hline
$\bar{S}$ & $-1$ & 1 & 1 & $-1$ & $-1$ \rule[-.2cm]{0cm}{.7cm} \\ \hline
$b_1$ & 1 & 1 & 1 & 1 & $ 1$ \rule[-.2cm]{0cm}{.7cm} \\ \hline
$b_2$ & 1 & 1 & 1 & $ 1$ & 1 \rule[-.2cm]{0cm}{.7cm} \\ \hline
\end{tabular}
\]
\centerline{Table 2.1: The coefficient $C_{(\alpha_i|\alpha_j)}$
is given by the $(i,j)$ entry of the matrix.}

This choice corresponds to a type IIA compactification\footnote{
The type IIA$\leftrightarrow$B exchange is realized by changing 
the chirality of, say, the right-moving spinors. In Appendix B we 
explain how this is implemented in the fermionic construction.}. 
Six of the eight gravitinos are projected out and we are left with 
only two supersymmetries, whose generators can be read-off from 
the $-1/2$ picture vertex operator representation of the surviving 
gravitinos, given in (\ref{gravi}). 

It is easy to check that the 
massless spectrum fits into representations of the $N=2$ supersymmetry.
This is done
by constructing the vertex operator representation of 
the states, which we quote in Appendix B.
The $N=2$ spectrum is in fact
characterized by the $SU(2)$ symmetry under which the two 
supercharges form a doublet. 
In Appendix B we discuss in detail the construction of the generators
of this $SU(2)$ symmetry.
By looking at the $SU(2)$ charge of the scalars, we 
identify the ones belonging to the vector multiplets and the ones belonging to 
the hypermultiplets: 
the scalars of a hypermultiplet do transform under the $SU(2)$ symmetry
of $N=2$ \cite{sugra}. In particular, it is easy to see that the pair 
dilaton--pseudoscalar is charged and therefore belongs to a hypermultiplet. 
Furthermore, it is also easy to see that all the scalars belonging to
hypermultiplets are charged also under a second $SU(2)$. 
This allows us to conclude that
the quaternionic manifold has an $SO(4)$ symmetry, and is 
given by the coset
\be
{ SO(4,4+N_H) \over SO(4) \times SO(4+N_H)}~,
\label{uth}
\ee
where $N_H$ is the number of hypermultiplets that originate
from the twisted sectors (in this case, these are the
sectors $b_1$, $b_2$ and $FS\bar{S}b_1b_2$, which,
for the choice of projections specified in Table 2.1,
provide the scalars of $N_H=12$ hypermultiplets\footnote{The scalars of 
the $S\bar{S}$ (Ramond--Ramond) sector
are charged also under two other $SU(2)$'s. They therefore have 
an $SO(4) \times SO(4)$ symmetry.
The four $SU(2)$'s are the remnant of the $SU(8)$ symmetry
of the massless spectrum of the $N=8$ theory, which is broken to $SU(2)^4$
by the orbifold projections.}.
A similar analysis, on the scalars uncharged under the
$SU(2)$ of the $N=2$ supersymmetry, allows us to conclude  
that the scalars belonging to vector multiplets span the coset:
\be
{SU(1,1) \over U(1)}
\times {SO(2,2+N_V) \over SO(2) \times SO(2+N_V)}, 
\label{utv}
\ee
with $N_V=12$ in this particular case.

We can construct other $Z_2 \times Z_2$ symmetric orbifolds, with 
a higher or lower number of massless states originating from the 
twisted sectors, by varying the sets $b_1$ and $b_2$ and/or adding 
more sets to the basis. However, it is easy to see that, once 
required that the breaking of the $N=8$ supersymmetry to $N=2$ be 
symmetric in the left- and right- movers, the untwisted sector is 
automatically fixed to be the same as for the orbifold considered 
above. By constructing then, for this class of orbifolds, the 
vertex operator representation of the massless states of the 
twisted sectors, it is easy to check that there is always an $SO(4) 
=SU(2) \times SU(2)$ symmetry common to all the scalars of the 
hypermultiplets. On the other hand, the complex scalars of the 
vector multiplets possess the $SO(2) \approx U(1)$ symmetry of 
complex conjugation, as the scalars of (\ref{utv}). As a 
consequence, the scalar manifolds are uniquely specified by the 
numbers $N_H$ and $N_V$ of hyper  and vector multiplets provided by 
the twisted sectors, and are always expressed by (\ref{uth}) and 
(\ref{utv}).

In order to give an exhaustive classification of such orbifolds,
we notice that, instead of varying the sets $b_1$, $b_2$, we
can equivalently keep them fixed and add to the fermion basis
the sets $e_i$:
\be
e_i=\left\{ y_i^L,\o_i^L~|~y_i^R,\o_i^R \right\}~~~~(i=1,...,5),
\ee
which factorize the six circles of the compact space
($e_6$ is generated by the product 
$FS\bar{S}$ $e_1$ $e_2$ $e_3$ $e_4$ $e_5$) by 
introducing independent $Z_2$ boundary conditions for all of them.

With such a basis, we can construct any $Z_2 \times Z_2$ orbifold, provided we 
properly choose the modular coefficients. In fact, with these fermion sets,
we can construct 48 massless twisted sectors\footnote{They are quoted 
in Appendix C.}, 
that is as many twisted sectors as the 
maximal number of fixed points 
a $Z_2 \times Z_2$ symmetric orbifold can have. 
Each such fixed point gives rise either to a vector- or to a hypermultiplet.
Any specific choice of the modular coefficients
amounts to a choice of GSO projections, which act by excluding
some sectors and determining whether the states of the
remaining sectors fit into vector- or hypermultiplets. 

We therefore proceed by expressing $N_V$ and $N_H$, for each twisted sector,
as functions of the modular coefficients (we quote the general
formula of the GSO projections on the 48 twisted sectors in Appendix C). 
Then we fix the coefficients that determine the GSO projections
onto the untwisted sector (RR sector included),
because they amount to an arbitrary choice of the chirality 
of the spinors; we then vary all the other GSO projections,
by allowing a change in the coefficients
$C_{(b_1|e_1)}$, $C_{(b_1|e_2)}$, 
$C_{(b_2|e_3)}$, $C_{(b_2|e_4)}$, $C_{(e_1|e_2)}$, $C_{(e_1|e_3)}$,
$C_{(e_1|e_4)}$, $C_{(e_1|e_5)}$, $C_{(e_2|e_3)}$, $C_{(e_2|e_4)}$,
$C_{(e_2|e_5)}$, $C_{(e_3|e_4)}$, $C_{(e_3|e_5)}$, $C_{(e_4|e_5)}$,
$C_{(b_1|Fe_3 e_4)}$, $C_{(b_2|Fe_1 e_2)}$ and $C_{(b_1 b_2 | e_5)}$.
In this way we obtain all the possible $(N_V,N_H)$ pairs.
The coefficient $C_{(b_1|b_2)}$ determines, instead, 
the general projection onto
the chirality of the bispinors of the twisted sectors. Under a change of sign
of this coefficient, $N_V$ and $N_H$ get exchanged. As a consequence,
each pair $(N_V,N_H)$ appears accompanied by its mirror $(N_H,N_V)$.
We list the pairs $(N_V,N_H)$ in Table D.1. 
Indeed, what we obtain is much more than a simple 
classification of the
possible massless spectra: having performed such an analysis on the single 
twisted sectors, we actually obtain a complete classification of the
possible orbifold projections, something that, as we will see in the following,
allows us to reconstruct the one-loop partition function of each orbifold,
even away from the fermionic point.

\noindent

\vskip 0.3cm
\setcounter{section}{3}
\setcounter{equation}{0}
\section*{\normalsize{\bf 3. The partition functions}}

In the type II $Z_2 \times Z_2$ symmetric constructions, the 
degrees of freedom of the compact space can be equivalently 
described by compactified bosons. 
The conditions of existence of the two world-sheet supercurrents
(\ref{TF}) allow in fact different $Z_2$
boundary conditions to be assigned not to single fermions but only to sets of 
bilinears of fermions.
The symmetry between left and right movers implies that such
bilinears must always appear paired in such a way as to form
compact bosons. Indeed, we want to show in the following that
all the above models can be constructed as orbifolds 
by using the symmetries of the conformal theory of six bosons
compactified on a torus $T^6$ at the point of moduli for which 
it is described by a product of circles
$T^6=S^1 \times \ldots \times S^1$
\footnote{In some cases, the factorization 
$T^6=T^2 \times T^2 \times T^2$ is sufficient.}.
In this approach, the dependence on the geometrical moduli of 
$T^6$ is explicit\footnote{In the fermionic construction  
the moduli dependence was not manifest, because compactified bosons
can be fermionized only for some particular values of moduli.
For instance, in the case of a single boson, the fermionic
partition function corresponds to the bosonic one when the radius of 
compactification $R$ is 1. 
The fermionic construction must, however, be considered 
as describing a model at a particular point in moduli space.}.
In order to see what is the partition function 
at a generic point in ${\cal T}^6$, we
use identities satisfied by the modular forms and recast 
the partition function of free fermions as a sum over lattice momenta
and windings, as in the case of a single boson.
By substituting generic values of moduli in the
lattice sums, wethen get the partition function of the model at any 
value in the orbifold moduli space.

The partition function, at the fermionic point (see Appendix A), 
is given by the integral over the modular
parameter $\tau$, with modular-invariant measure 
$(\Im \, \tau)^{-2} d \tau d \bar{\tau} $, of:
\be
Z^{{\rm string}} =   {1 \over \Im \tau | \eta(\tau)|^4}~  
{1 \over 4} \sum_{(H_1,G_1,H_2,G_2)}~
\left( {1 \over 2} \right)^6  \sum_{(\gamma,e_i,\delta,d_i)}
C {\tiny \ar{\gamma,e_i,H_j}{\delta,d_i,G_j} }
~Z^F_{\rm L}~  Z^F_{\rm R} ~Z_{6,6}~,
\label{z}
\ee
where $Z^F_{L,R}$ contain the contribution of the world-sheet 
fields $\psi_{\mu}^{L,R}$, $\c_{a}^{L,R}$ (the sets $S$ and $\bar{S}$); 
$Z_{6,6}$ encodes the contribution of the $c=(6,6)$ internal space, 
i.e. of the fields $\o_I^{L,R}$, $y_I^{L,R}$ (the fields of the sets
$\Gamma \equiv FS\bar{S}$, $e_i$, $i=1,...,5$ and their products) and 
$C \ar{\gamma,e_i,h_j}{\delta,d_i,g_j}$ is a modular covariant phase
(discrete torsion).
We have:
\be
Z^F_{\rm L}={1 \over 2} \sum_{(a,b)} {{\rm e}^{i \pi \varphi_{\rm L} 
\left( a,b,\vec H,\vec G \right)} \over \eta^4}
\vartheta {\tiny \ar{a}{b}}\vartheta {\tiny \ar{a+H_1}{b+G_1}}
\vartheta {\tiny \ar{a+H_2}{b+G_2}}\vartheta {\tiny \ar{a-H_1-H_2}{b-G_1-G_2}}
~, \label{fl}
\ee
\be
Z^F_{\rm R}={1 \over 2} \sum_{(\bar{a},\bar{b})} {{\rm e}^{i \pi \varphi_{\rm R} 
\left( \bar{a},\bar{b}, \vec H, \vec G \right)}
 \over \bar{\eta}^4}
\vartheta {\tiny \ar{\bar{a}}{\bar{b}}}\vartheta {\tiny \ar{\bar{a}+H_1}
{\bar{b}+G_1}}
\vartheta {\tiny \ar{\bar{a}+H_2}{\bar{b}+G_2}}\vartheta {\tiny \ar{\bar{a}-
H_1-H_2}{\bar{b}-G_1-G_2}}~,
\label{fr}
\ee
with
\ba
\varphi_{\rm L} 
\left( a,b, \vec H, \vec G \right)& = &
a+b+{1 \over 2}\left( 1-C_{(S|S)} \right) a b
+{1 \over 2}\left( 1-C_{(S|S\bar{S}b_1)}\right) 
\left( a G_1+b H_1 \right)\nn \\
&&+{1 \over 2}\left( 1-C_{(S|S\bar{S}b_2)}\right) 
\left( a G_2+b H_2 \right)
\ea
and an analogous expression for
$\varphi_{\rm R} 
\left( \bar{a},\bar{b}, \vec H, \vec G \right)$,
obtained from $\varphi_{\rm L}$ through the substitutions 
$(a,b) \to (\bar{a},\bar{b})$ and $S \to \bar{S}$.
The contribution of the compact bosons is:
\ba
Z_{6,6} & = & 
{1 \over  |\eta|^4}\, \left\vert \vartheta \ar{\gamma+e_1}{\delta+d_1}
\vartheta \ar{\gamma+e_1+H_2}{\delta+d_1+G_2} 
\vartheta \ar{\gamma+e_2}{\delta+d_2}
\vartheta \ar{\gamma+e_2+H_2}{\delta+d_2+G_2} \right\vert 
 \nonumber \\ && \nonumber \\
& \times & 
{1\over |\eta|^4}\, \left\vert \vartheta \ar{\gamma+e_3}{\delta+d_3}
\vartheta \ar{\gamma+e_3+H_1}{\delta+d_3+G_1}
\vartheta \ar{\gamma+e_4}{\delta+d_4}
\vartheta \ar{\gamma+e_4+H_1}{\delta+d_4+G_1} \right\vert
 \label{zb}\\ && \nonumber \\
& \times &
{1\over |\eta|^4}\, \left\vert \vartheta \ar{\gamma+e_5}{\delta+d_5}
\vartheta \ar{\gamma+e_5+H_1+H_2}{\delta+d_5+G_1+G_2}
\vartheta \ar{\gamma}{\delta}
\vartheta \ar{\gamma+H_1+H_2}{\delta+G_1+G_2} \right\vert~. \nonumber
\ea
In this notation, the pairs $(a,b)$ and $(\bar{a},\bar{b})$ specify the
boundary conditions, in the directions ${\bf 1}$ and $\tau$ of the
world-sheet torus, of the sets $S$ and $\bar{S}$, while
$(\gamma,\delta)$, $(e_i,d_i)$ refer respectively to the sets $\Gamma$ and 
$e_i$; $(H_1,G_1)$ and $(H_2,G_2)$ refer to the sets $b_1$, $b_2$.
When a field belongs to the intersection of many sets, its boundary conditions
are specified by the sum of the boundary conditions of the sets it belongs to.
The modular coefficients appear in the phases 
$\varphi_{\rm L}$, $\varphi_{\rm R}$ in $Z^F_{L,R}$
and in 
\be
C \ar{\gamma,e_i,h_j}{\delta,d_i,g_j}=
\exp~ {i \pi  \over 2} \sum_{k,\ell} \left(1 - C_{(X_k|X_{\ell})} \right)
\a_k \b_{\ell}~,
\label{cx}
\ee
where
\be
X_k,\,X_{\ell}~ \in~ \left\{\Gamma, b_1,b_2, e_i \right\}
\ee
and $(\a_k, \b_{\ell})$ indicate the corresponding boundary conditions
in the two directions of the world-sheet torus. 
For the specific choice of Table 2.1 (type IIA), we have
\ba
\varphi_{\rm L} & = & a+b+a b ~, \\
\varphi_{\rm R} & = & \bar{a}+\bar{b}+\bar{a} \bar{b}~.
\ea
For the type IIB choice specified in Appendix B.2, $\varphi_{\rm R}$ is,
instead:
\be
\varphi_{\rm R}  =  \bar{a}+\bar{b}~.
\ee
The partition function is the sum of five terms: 
\begin{enumerate}
\item the $N=8$ sector, specified by
$(H_1,G_1)=(H_2,G_2)=(0,0)$; 
\item the $N=4$ sector specified by
$(H_1,G_1) \neq (0,0)$, $(H_2,G_2) = (0,0)$; 
\item the $N=4$ sector with
$(H_2,G_2) \neq (0,0)$, $(H_1,G_1) = (0,0)$; 
\item the $N=4$ sector with
$(H_1,G_1) = (H_2,G_2) \neq (0,0) $ and
$(H_1+H_2,G_1+G_2) = (0,0)$;
\item the $N=2$ sector,
which contains all the terms for which $(H_1,G_1) \neq (0,0)$, 
$(H_2,G_2) \neq (0,0)$, $(H_1+H_2,G_1+G_2) \neq (0,0)$. 
\end{enumerate}
The $N=8$ sector is universal: it is the same for any orbifold,
since it is proportional to the unprojected partition function
of the $N=8$ string. In the $N=2=(1,1)$ sector all the bosons of
the compact space are twisted and/or projected: this implies that the
part of the
partition function that corresponds to this sector is the same at any
point in the moduli space of the orbifold.
The only non-trivial moduli dependence is contained in the $N=4$ sectors:
in the following, we will therefore concentrate on these.

In each $N=4$ sector the moduli dependence is contained in the 
untwisted  $c=(2,2)$ conformal block. The latter corresponds to
the complex planes (1,2) (for the first $N=4$ sector),  
(3,4) in the second sector 
and (5,6) in the third sector.
We want to rewrite such blocks in terms of sums over
lattice windings and momenta. To this purpose, 
we make use of the identity:
\be
\Gamma_{2,2}^{w}\ar{h_1,\;h_2}{g_1,\;g_2}(T(w),U(w))=\sum_{a_1,b_1,a_2,b_2}
{\rm e}^{i \pi (a_1g_1+b_1h_1+h_1g_1)} {\rm e}^{i \pi (a_2g_2+b_2h_2+h_2g_2)}
\left| \vartheta \ar{a_1}{b_1}\vartheta \ar{a_2}{b_2} \right|^2~,
\label{gs}
\ee
which generalizes the equivalence of the partition functions of two 
Weyl--Majorana fermions and one boson at radius 1 to the case of two bosons
toroidally compactified, with generic lattice shifts $(h_1,h_2,g_1,g_2)$
in the momenta and windings in the two circles. Here $w \equiv (w_1,w_2)$ 
stays for a pair of lattice vectors $w_1$, $w_2$, which specify the
directions of the shifts (see Appendix E). We do not
need to specify the particular value, which depends
on the shift vectors, of the toroidal K{\"a}hler and 
complex structure moduli $T$ and $U$ for which the equivalence (\ref{gs}) is 
valid. This, however, can be easily computed, and we refer to Appendix E
for this detail.
In our case, the pairs $(a_1,b_1)$, $(a_2,b_2)$ are substituted by
$(\gamma+e_1,\delta+d_1)$, $(\gamma+e_2,\delta+d_2)$
in the first $N=4$ sector, 
$(\gamma+e_3,\delta+d_3)$, $(\gamma+e_4,\delta+d_4)$ in the second,
and $(\gamma+e_5,\delta+d_5)$, $(\gamma+e_6,\delta+d_6)$ in the third.

The shifts $(h_1,g_1)$, $(h_2,g_2)$, for any specific case, are 
read-off from (\ref{z}) and (\ref{cx}). Owing to the fact that in 
a twisted/shifted lattice character $\Gamma \ar{h|h'}{g|g'}$, when 
$(h,g) \neq (0,0)$ the twist $(h,g)$ imposes a constraint on the 
shift $(h',g')$ ($(h',g')=(0,0)$ or $(h',g')=(h,g)$), it turns out 
that in the $N=4$ sectors all the shifts can be expressed in terms 
of the two $Z_2$ supersymmetry-breaking projections introduced by 
the sets $b_1$ and $b_2$ (these may or may not act freely, by 
translating some of the coordinates of the compact space), and in 
terms of the projections associated to the symmetries of each 
$c=(1,1)$, $S^1/Z_2$ orbifold, generated by the two elements 
\cite{dvv}: 
\ba D:~~~~(\sigma_+,\sigma_-,V_{nm})& \to & 
(\sigma_-,\sigma_+,(-)^m V_{nm}), \label{d} \\ 
\tilde{D}:~~~~(\sigma_+,\sigma_-,V_{nm})& \to & 
(-\sigma_+,\sigma_-, (-)^n V_{nm}). 
\ea 
Here $\sigma_+$, 
$\sigma_-$ are the two twist fields of $S^1/Z_2$, and $V_{nm}$ are 
the untwisted vacua labelled by the momentum $m$ and the winding 
number $n$. The orbifolds we are considering indeed possess such 
symmetries. In fact, the presence, in the fermionic basis, of the 
sets $e_i$, $i=1,\ldots,5$ corresponds to a choice of coordinates 
for which the compact space is described by a product of circles: 
$T^6=S^1 \times \ldots \times S^1$. Any one of the $Z_2\left( b_i 
\right)$ projections ($i=1,2,3=1+2$), then creates a $c=(4,4)$ 
twisted block that corresponds to an orbifold $\left( S^1 
\right)^4 /Z_2 \left( b_i \right)$. (The construction with complex 
planes corresponds instead to the separation $\left[ T^2 \times 
T^2 \right] / Z_2 \left( b_i \right)$. At this point in the moduli 
space, it is possible to remove some of the fixed points, by using 
such symmetries. To simplify the discussion, we 
will consider in the following only the first generator, $D$ (the action of 
$\tilde{D}$ can be obtained by $T$-duality). Modding out by the 
group generated by $D$ amounts to cutting half of the states in 
the twisted sector and to a modification of the lattice of momenta 
and windings: the momenta are restricted to even values and 
the windings are shifted to half-integer values. In order to 
realize this projection, we must pair the $D$-projection on the 
twisted $c=(4,4)$ block with a translation in one direction of the 
untwisted $c=(2,2)$ block. From (\ref{d}), we see that
this translation can itself be considered as a $D$-projection.
The operation therefore amounts to the insertion of a $D$-projection
into two circles belonging to two
different complex planes. There are then always two $N=4$ sectors for
which the pair of $D$-operations acts by reducing the number of fixed points.

In order to account for the various possibilities, we 
extend the definition of
twisted/shifted $c=(2,2)$ conformal blocks to account also for the 
$D$-operations. We therefore define 
\be
\Gamma_{2,2}^{(i)} \ar{H;\;h_1,\;h_2~|~H^1,\;H^2}
{G;\;g_1,\;g_2~\,|~G^1,\;G^2},
~~~~~~~i=1,2,3,
\label{gts}
\ee
where the index $i$ indicates the planes (1,2), (3,4) and (5,6) 
respectively. The pair $(H,G)$ specifies the 
ordinary twist, the pairs $(h_1,g_1)$, $(h_2,g_2)$
specify the lattice shifts in the two circles of $T^2$,
while the pairs $(H^1,G^1)$, $(H^2,G^2)$
refer to the $D$-operations. When $(H,G) \neq (0,0)$, the block
is non-zero only if $(H^1,G^1)$, $(H^2,G^2)$ equal $(0,0)$ or $(H,G)$.
In this case, also the shifts $(h_1,g_1)$, $(h_2,g_2)$ 
are constrained in the same way: $(h_1,g_1)$, $(h_2,g_2)=(0,0)$
or $(H,G)$ and we set, by definition, 
\be
\Gamma_{2,2} \ar{H;\;h_1,\;h_2~|~H^1,\;H^2}
{G;\;g_1,\;g_2~\,|~G^1,\;G^2}_{(H,G) \neq (0,0)} \equiv
 {4 \left\vert \eta \right\vert^6 \over 
\left\vert  \vartheta\ar{1+H}{1+G} \vartheta\ar{1-H}{1-G}            
\right\vert }~,
\ee 
\be
(h_i,g_i),~(H^j,G^j)=(0,0)~~{\rm or}~~=(H,G)~. 
\nonumber
\ee
When $(H,G)=(0,0)$, the $D$-action amounts to a shift, which adds to the
shifts $(h_1,g_1)$, $(h_2,g_2)$. In this case,
(\ref{gts}) is a doubly shifted lattice sum; the lattice shifts in the
two circles are specified by $(h_1+H^1,g_1+G^1)$
and  $(h_2+H^2,g_2+G^2)$:
\be
\Gamma_{2,2}\ar{0;\;h_1,\;h_2~|~H^1,\;H^2}
{0;\;g_1,\;g_2~\,|~G^1,\;G^2} =
\Gamma_{2,2} \ar{h_1+H^1,h_2+H^2}{g_1+G^1,g_2+G^2}~.
\ee
There is, however, a subtlety:
the pair of $D$-operations was defined as
a projection, which mods out states by using a symmetry of the twisted sector.
It makes sense only
when the $Z_2 \left( b \right)$ projections do not act freely.
The recipe is that one first 
projects with at least one $Z_2\left( b \right)$, 
to reduce supersymmetry to $N=4$ or $N=2$,
then the $D$-projection can be inserted in the 
$Z_2\left( b \right)$-twisted sectors
and related to a shift in the untwisted coordinates. 
If the $Z_2\left( b \right)$ acts freely, i.e. has no fixed points, 
such an operation cannot be performed. 
This means that the $D$-operation is not independent
of the shifts $(h_i,g_i)$, $i=1,2$, which in turn depend on the two 
projections introduced by $b_1$ and $b_2$. We can however extend
the definition of the $D$-projection to include also the case of freely acting
orbifolds, by specifying that, in the absence of fixed points, it acts simply
as a shift, i.e. as the natural restriction of (\ref{d}), 
with the constraint that it must always act in
a direction independent of that of $(\vec{h},\vec{g})=
\left\{ (h_1,g_1),(h_2,g_2) \right\}$.
Its interference with the shift $(\vec{h},\vec{g})$ then produces the
following shifted lattice sum:
\be
{1 \over 2} \Gamma^{w_1}_{2,2} \ar{H}{G}+ {1 \over 2} 
\Gamma^{w_2}_{2,2} \ar{H}{G}~, ~~~~\left( w_1 \right)^2=\left( w_2 
\right)^2=w_1 \cdot w_2=0~. 
\ee 
Once this is pointed out, we can 
unambiguously express, in full generality, the $c=(6,6)$ conformal 
block $Z_{6,6}^{\vec{T},\vec{U}}$ of the orbifold partition 
function at a generic point in the space of the K{\"a}hler and the complex 
structure moduli, $\vec{T} \equiv (T^1,T^2,T^3)$ and $\vec{U} 
\equiv(U^1,U^2,U^3)$, as a product of the above defined 
twisted/shifted characters: 
\be
Z_{6,6}^{\vec{T},\vec{U}} 
 =   {1 \over \left\vert \eta   \right\vert^{12}}
\Gamma_{6,6}
\ar{H_1,\;H_2,~\vec{h}~|\;\vec{H}}{G_1,\;G_2,~\vec{g}~|\;\vec{G}}
 \equiv  
Z_{6,6}^{\vec{T},\vec{U}} 
\ar{H_1,\;H_2,~\vec{h}~|\;\vec{H}}{G_1,\;G_2,~\vec{g}~|\;\vec{G}}~,
\label{z6ts1}
\ee
where
\ba
\Gamma_{6,6}
\ar{H_1,\;H_2,~\vec{h}~|\;\vec{H}}{G_1,\;G_2,~\vec{g}~|\;\vec{G}}
& = &
\Gamma_{2,2}^{(1)} \ar{H_2,\vec{h}_{(1)}~|\;H^1,\;H^{2}}
{G_2,\vec{g}_{(1)}~|\;G^1,\;G^{2}}  \times
\Gamma_{2,2}^{(2)} \ar{H_1,\vec{h}_{(2)}~|\;H^3,\;H^{4}}{G_1,
\vec{g}_{(2)}~|\;G^3,\;G^4} \nn \\
&& \times \,
\Gamma_{2,2}^{(3)} \ar{H_1+H_2,\vec{h}_{(3)}~|\;H^5,\;H^{6}}
{G_1+G_2,\vec{g}_{(3)}~|\;G^5,\;G^{6}}~.
\ea
In the above expression,
the shifts $(\vec{h},\vec{g})$ depend on the projections
$Z_2 \left( b_i \right)$, and can always be expressed in terms
of the twists
$(H_1,G_1)$, $(H_2,G_2)$, while $\vec{H} \equiv (H^1,\ldots,H^6)$, 
$\vec{G}\equiv (G^1,\ldots,G^6)$ refer to the $D$-projections.
At a generic point in the moduli space, the partition function (\ref{z})
becomes:
\be
Z^{{\rm string}}(\vec{T},\vec{U}) =   {1 \over \Im \tau | \eta(\tau)|^4}~  
{1 \over 4} \sum_{(H_1,G_1,H_2,G_2)}
\left( {1 \over 2} \right)^{n_D}  \sum_{(\vec{H},\vec{G})}
~C \ar{H_1,~H_2,~H^i,~H^j}{G_1,~G_2,~G^i,~G^j}
Z^F_{\rm L}~  Z^F_{\rm R} ~Z_{6,6}^{\vec{T},\vec{U}}~. 
\label{ztu}
\ee
$Z_{L,R}^F$ are defined as in (\ref{fl}), (\ref{fr});
$n_D$ indicates the number of $D$-projections. Notice that even though
$n_D$ can be greater than 2, in each $N=4$ sector the maximal number
of projections that effectively act is 2, because there are only
two independent directions in which it is possible to pick a 
modular-invariant shift (see Appendix C of \cite{6auth}).
Further projections superpose and their effect vanish.
What remains of the coefficient
$C {\tiny \ar{\gamma,e_i,H_j}{\delta,d_i,G_j} }$ of expression (\ref{z}) is:
\be 
C \ar{H_1,~H_2,~H^i,~H^j}{G_1,~G_2,~G^i,~G^j}=
{\rm e}^{ {i \pi  \over 2} \left(1-C_{(b_1|b_2)} \right)
\left(H_1G_2+H_2G_1 \right) }
\prod_{ij}C_{ij}\ar{H^i,~H^j}{G^i,~G^j},
\label{C}
\ee
where $C_{ij}$ can be either $+1$ or $(-)^{H^iG^j+H^jG^i}$; $(H^i,G^i)$,
$(H^j,G^j)$ refer to the $D$-operations in the circles $i$, $j$.
These coefficients are always $+1$ in the $N=4$ sectors; however they
can play a role in the $N=2$ sector.

As is clear from the definition of twisted/shifted lattice given in
(\ref{gts}) and its properties, 
an $N=4$ sector can provide 16, 8, 4 or 0 supermultiplets,
depending on the shift $(\vec{h},\vec{g})$ and on the $D$-projections.
Thanks to the interpretation of the different choices of modular
coefficients in terms of such operations, we understand why
it is not possible to obtain models with any number 
$N_V+N_H$ of supermultiplets between 48 and 0, modulo 4 (with the obvious
exception of 44, which would require one $N=4$ sector with twelve
supermultiplets). We saw that,
in order to effectively reduce the number of states, the 
$D$-operation must always be inserted in at least two
circles belonging to two different complex planes, which  implies that
there are always at least two $N=4$ sectors in which eight supermultiplets
become massive. As a consequence, the maximal number $N_V+N_H$ of 
supermultiplets originating from the twisted sectors, below the 48,
which is reached only when all the shifts $\vec{h},\vec{g}$ and the 
$D$-projections are turned off (all the modular coefficients are $+1$), 
is 32, and is obtained when $n_D=1$ and
the $D$-projection involves only two different complex planes.
This explains why, in the classification
of Table D.1, there are no models with $N_V+N_H$ between 48 and 32.
For an analogous reason, also $N_V+N_H=28$ is forbidden: this would require
one more $D$-projection, reducing further the number of states
in only one  $N=4$ sector, but it is easy to realize that in order to reduce
the number of supermultiplets to four, the two $D$-projections must be
inserted in at least four different circles, two per projection.
There are therefore always at least two $N=4$ sectors with 4
supermultiplets, so that, below 32, the maximal number of
supermultiplets is 24. Below this, all the numbers modulo 4 are allowed:
it is possible to construct models with $N_V+N_H=20, 16, 12,
8, 4, 0$. One can check that these are precisely the numbers that appear
in the list of Table D.1. 
In Section 4.2 we will also see
how the operations described above can be interpreted in terms of 
stringy (super-)Higgs mechanisms.

\noindent

\vskip 0.3cm
\setcounter{section}{4}
\setcounter{equation}{0}
\section*{\normalsize{\bf 4. Helicity supertraces and
(super-)Higgs phenomena }}

When specified for a certain model, 
formula (\ref{ztu}) encodes, in principle, all the perturbative information
about it. It can be used to investigate the BPS spectrum
and to compute one-loop threshold corrections.
In each model, all the non-trivial moduli dependence is contained in the
three $N=4$ sectors, so that essentially the classification of 
$Z_2 \times Z_2$ symmetric orbifolds amounts to
assigning the three $N=4$ sectors for
each one of the massless spectra appearing in Table D.1.
According to the analysis of Section 3, this is equivalent to
specifying the form of the lattice sum, $\Gamma_{2,2}^{(i)}$,
for each one of the three untwisted tori.
The set $(N_V,N_H,\Gamma_{2,2}^1,\Gamma_{2,2}^2,\Gamma_{2,2}^3)$
then fixes unambiguously the entire partition function.  
Actually, as far as we are interested only in the
$N=4$ sectors, the notation (\ref{gts}) is highly
redundant, because the shifts and the $D$-projections
are constrained, in each $(H,G)$-twisted sector, to be
either $(0,0)$ or equal to $(H,G)$. 
It is therefore sufficient to specify the direction
and the nature of the translations through a pair of lattice vectors,
$w_1$ and $w_2$ (see Appendix E).  
We can then account for the various situations by
introducing the following notation for the $(H,G)$-shifted lattice sums:
\ba
O &  = &  \Gamma_{2,2}\ar{0}{0}~,
\label{O} \\ 
&& \nonumber \\ 
F & = & \Gamma_{2,2}^{w_1}\ar{H}{G}~, 
\label{F}\\
&& \nonumber \\ 
FD
&=& {1 \over 2}\Gamma_{2,2}^{w_1}\ar{H}{G}~+
{1 \over 2}\Gamma_{2,2}^{w_2}\ar{H}{G}~ ,
\label{FD}\\
&&\nonumber\\
D &  = &  {1 \over 2}\Gamma_{2,2}\ar{0}{0}~+
                           {1 \over 2}\Gamma^{w_1}_{2,2}\ar{H}{G}~ , 
\label{D}\\
&& \nonumber \\
DD &  = &  
     {1 \over 4} \Gamma_{2,2} \ar{0}{0}~+
     {1 \over 4}\Gamma_{2,2}^{w_1}\ar{H}{G}~+ \nonumber \\
&&     +{1 \over 4}\Gamma_{2,2}^{w_2}\ar{H}{G}~+
     {1 \over 4}\Gamma_{2,2}^{w_1+w_2}\ar{H}{G}~ . 
\label{DD}
\ea
The shift vectors satisfy:
\be
\left( w_1 \right)^2=\left( w_2 \right)^2=w_1 \cdot w_2=0
\ee
and we indicated by $\Gamma_{2,2} \ar{0}{0}$ the ordinary
unshifted lattice sum. The result of our analysis,
which accounts for all the possible
``partition functions'', is quoted in Appendix C.

\subsection*{\normalsize{\sl 4.1. Helicity supertraces}}

We are interested in the second and fourth helicity supertraces, $B_2$ and 
$B_4$, through which we 
control the behaviour of a model under a motion in moduli space.
The helicity supertraces are defined, 
for a given representation $R$ of supersymmetry, as
\be
B_{2n}(R) \equiv \hbox{Str} \lambda ^{2n}= 
\hbox{Tr}_{\rm R}[(-)^{2 \lambda} \lambda ^{2n}],
\ee
where $\lambda$ stands for the physical four-dimensional helicity.
In the framework of string theory, $\lambda =\lambda_{\rm L}+\lambda_{\rm R}$,
where $\lambda_{\rm L,R}$ are the contributions to the helicity from the
left and right movers. The quantities $B_{2n}$ are computed by
taking appropriate derivatives of the generating function
\be
Z^{{\rm string}}(v,\bar{v})=\hbox{Tr}'q^{L_0-{c \over 24}}
\bar{q}^{\bar{L}_0-{\bar{c} \over 24}} {\rm e}^{2 \pi i (v\lambda_{\rm L}-
\bar{v}\lambda_{\rm R})}~,
\ee 
by defining
\be
\lambda_{\rm L}=Q={1 \over 2 \pi i} {\p \over \p v},~~~~
\lambda_{\rm R}=\bar{Q}={1 \over 2 \pi i} {\p \over \p \bar{v}}.
\ee
We then have 
\be
B_{2n}^{{\rm string}}=
(Q+\bar{Q})^{2n} Z^{{\rm string}}(v,\bar{v}) \vert _{(v=\bar{v}=0)}.
\ee
An explicit expression for $Z^{{\rm string}}(v,\bar{v})$, in the case of 
$Z_2 \times Z_2$ symmetric orbifolds, 
is given by an expression that is similar to the
$v=\bar{v}=0$ case presented in (\ref{ztu}):
\be
Z^{{\rm string}}(v,\bar{v}) =   {1 \over \Im \tau | \eta(\tau)|^2}~  
{1 \over 4} \sum_{(H_1,G_1,H_2,G_2)}~
\left( {1 \over 2} \right)^{n_D}  \sum_{(\vec{H},\vec{G})}
~Z^F_{\rm L}(v)~  Z^F_{\rm R}(\bar{v})~\xi(v) \bar{\xi}(\bar{v}) 
~Z_{6,6}^{\vec{T},\vec{U}}~, 
\label{zv}
\ee
where 
\be
\xi(v)=\prod_{n=1}^{\infty} {(1-q^n)^2 \over (1-q^n {\rm e}^{2 \pi i v})
(1-q^n {\rm e}^{-2 \pi i v})}=
{\sin \pi v \over \pi} {\vartheta'_1(0) \over \vartheta_1(v)}
\label{xi}
\ee
is an even function of $v$ ($\xi(v)=\xi(-v)$) that 
counts the helicity contributions of the space-time bosonic
oscillators and $Z^F_{\rm L}(v)$, $Z^F_{\rm R}(\bar{v})$ are 
the contributions of the world-sheet fields $\psi^L_{\mu}$, $\c^L_I$,
$\psi^R_{\mu}$, $\c^R_I$, (\ref{fl}) and (\ref{fr}), 
modified by a change in the argument of the theta functions:
\be
Z^F_{\rm L}(v)={1 \over 2} \sum_{(a,b)} {{\rm e}^{i \pi \varphi_{\rm L}} \over 
\eta^4}~
\vartheta {\tiny \ar{a}{b}}(v)~\vartheta {\tiny \ar{a+H_1}{b+G_1}}
\vartheta {\tiny \ar{a+H_2}{b+G_2}}\vartheta {\tiny \ar{a-H_1-H_2}{b-G_1-G_2}}
~, \label{flv}
\ee
and
\be
Z^F_{\rm R}(\bar{v})={1 \over 2} \sum_{(\bar{a},\bar{b})} 
{ {\rm e}^{i \pi \varphi_{\rm R}}
 \over \bar{\eta}^4}~
\bar{\vartheta} {\tiny \ar{\bar{a}}{\bar{b}}}(\bar{v})~
\bar{\vartheta} {\tiny \ar{\bar{a}+H_1}{\bar{b}+G_1}}
\bar{\vartheta} {\tiny \ar{\bar{a}+H_2}{\bar{b}+G_2}}
\bar{\vartheta} {\tiny \ar{\bar{a}-
H_1-H_2}{\bar{b}-G_1-G_2}}~,
\label{frv}
\ee
In order to compute the quantities $B_{2n}$, we observe that, by
using the Riemann identity, these two terms can be cast in the form
\ba
Z^F_{\rm L}(v) & = & {1 \over \eta^4}~
\vartheta \ar{1}{1}\left({v \over 2}\right)~
\vartheta \ar{1+H_1}{1+G_1}\left({v \over 2}\right)~
\vartheta \ar{1+H_2}{1+G_2}\left({v \over 2}\right)~
\vartheta \ar{1-H_1-H_2}{1-G_1-G_2}\left({v \over 2}\right)
~, \nonumber \\
Z^F_{\rm R}(\bar{v}) & = & {1 \over \bar{\eta}^4}~
\bar{\vartheta} \ar{1}{1}\left({\bar{v} \over 2}\right)~
\bar{\vartheta} \ar{1+H_1}{1+G_1}\left({\bar{v} \over 2}\right)~
\bar{\vartheta} \ar{1+H_2}{1+G_2}\left({\bar{v} \over 2}\right)~
\bar{\vartheta} 
\ar{1- H_1-H_2}{1-G_1-G_2}\left({\bar{v} \over 2}\right)~.
\nn \\
&&
\ea
Then, to evaluate the various derivative terms, we use the properties that 
$\vartheta \ar{1}{1}$ and its even derivatives with respect to $v$
are odd under $v \to -v$ and vanish at $v=0$.

Taking this into account, it is easy to see that the only non-zero
contribution to the second helicity supertrace, $B_2$:
\be
B_2=(Q+\bar{Q})^2~Z^{{\rm string}} (v,\bar{v}) \vert_{v=\bar{v}=0}~,
\label{b2q}
\ee
comes from the $N=2$ sector. This is easily computed to be a constant:
\ba
B_2 & = & {1 \over 2 |\eta|^{12}} \Re \sum_{H_1,G_1,H_2,G_2}
\left({1 \over 2} \right)^{n_D} \sum_{\vec{H},\vec{G}}
C\ar{H_1,~H_2, ~H^i,~H^j}{G_1,~G_2, ~G^i,~G^j}
Z_{6,6}^{\vec{T},\vec{U}} \nonumber \\
& = & 8 \sum_{H_1,G_1,H_2,G_2}
\left({1 \over 2} \right)^{n_D} \sum_{\vec{H},\vec{G}}
C\ar{H_1,~H_2, ~H^i,~H^j}{G_1,~G_2, ~G^i,~G^j} \label{b2c} \\
&& \nonumber \\
& = & N_V-N_H, \nonumber
\ea
in agreement with the supergravity
computation, for which the gravity multiplet, as well
as a vector multiplet, contribute $+1$, while a hypermultiplet and a 
hypertensor multiplet
(the multiplet that contains the dilaton) contribute $-1$. In this way
the contribution of the untwisted sector cancels, leaving precisely
the difference between the number of vector and hypermultiplets
coming from the twisted sectors. The result (\ref{b2c}) also shows that,
for the symmetric constructions we are considering, the contribution
of the massive short multiplets sums up to zero\footnote{The 
contribution of a short massive multiplet $S^j$ of spin $j$ is 
$B_2(S^j)=(-)^{2j+1}(2j+1)$, so that (\ref{b2c}) puts constraints on the ratios
of the numbers of massive short multiplets of integer and half-integer spin.}.

It is also easy to see that $B_4$, which 
counts the number of short multiplets,
\be
B_4=(Q+\bar{Q})^4~Z^{{\rm string}} (v,\bar{v}) \vert_{v=\bar{v}=0}~,
\label{b4q}
\ee
receives contributions from both the $N=2$ and the
$N=4$ sectors. The contribution of the $N=2$ sector is due to
the term $4(Q^3 \bar{Q}+Q \bar{Q}^3)~Z^{{\rm string}} (v,\bar{v}) 
\vert_{v=\bar{v}=0}$ in the expansion of (\ref{b4q}), and 
in the models we are considering it turns out to be
equal to $B_2$, while the contribution 
of the $N=4$ sectors is due to the term $6 Q^2 \bar{Q}^2~ 
Z^{{\rm string}} (v,\bar{v}) \vert_{v=\bar{v}=0}$.
The contribution of each $N=4$ sector is easily computed:
\be
B_4^{(i)}=6 |\eta|^4 \Sump Z_{2,2}^{(i)}(X) \ar{H}{G}~,
\label{b4}
\ee
where the prime on the summation means that the value $(H,G)=(0,0)$ is
excluded and $Z_{2,2}^{(i)}(X) \ar{H}{G}$ is the conformal block 
that encodes the contribution of the $i$-th unshifted plane:
\be
Z_{2,2}^{(i)}(X) \ar{H}{G}={ X \over |\eta|^4},
\label{b44}
\ee
where $X$ stands for one of the expressions (\ref{O})--(\ref{DD}).
The massless limit of (\ref{b44}) is:
\be
B_4^{(i)} \limit{\longrightarrow}{\Im \tau \to \infty}
6+{3 \over 4}\left( N_V^{(i)}+N_H^{(i)} \right) \, ,
\ee
where $N_V^{(i)}$ and $N_H^{(i)}$ are respectively the number
of vector multiplets and hypermultiplets originating from the
$i$-th twisted sector.
Summing over the three sectors
and adding the contribution of the $N=2$ sector, we
obtain the expected massless contribution,
in agreement with supergravity:
\be
B_4 \vert_{\rm massless}=18+ { 7 N_V-N_H \over 4}~. 
\label{B4m}
\ee
Since the threshold corrections in general are expressed in terms of
integrals over the fundamental domain of torus partition functions,
for later convenience we quote in Appendix E also the integrals
of the various $Z_{2,2}(X)$.

\subsection*{\normalsize{\sl 4.2. Higgs and super-Higgs phenomena}}

It is a general property of
shifted lattices, $\Gamma_{2,2}^w \ar{H}{G}$, that there is
always at least one corner in moduli space in which, for
$(H,G)=(1,0)$ or $(1,1)$, the lattice sum vanishes
(for a detailed account, see for instance \cite{kkprn}). The particular
limit(s) at which this happens depends on the shift vector $w$,
and, once specified the modular properties of the lattice sum,
the various situations, which differ in the choice of $w$,
are mapped into one another by $SL(2,Z)$ transformations performed on
the toroidal moduli $T$ and $U$ (see for instance \cite{kkprn}).
The vanishing of the lattice sum at a particular limit
means that the states originating from the $(H,G)$-twisted sector 
become infinitely massive and decouple from the spectrum.
In this limit the unprojected theory is recovered.

This phenomenon 
is reflected in the behaviour of $B_4$, as it appears from
eq. (\ref{b4}). The contribution of a given $N=4$ sector
vanishes completely in an appropriate limit in the $(T,U)$ space
when  $X$ is given by expression
(\ref{F}) or (\ref{FD}), namely when the $Z_2$ projection
which breaks supersymmetry acts freely in that sector. 
In the cases (\ref{D}) and (\ref{DD}), instead, there is
a decoupling of respectively one-half and three-quarters
of the states of the twisted sector. 
In some cases, in the limit in which 
the states with shifted mass decouple from the spectrum,
there is also an  associated effective 
restoration of a certain number of supersymmetries \cite{kk,solving}.
In this case, the decoupling of some states
is accompanied by the appearance of new massless states,
which fit into multiplets of the enlarged supersymmetry.
This happens in the limits in which one or both of the $Z_2(b)$ projections 
effectively vanish.
We can therefore restore four or even eight supersymmetries.
A necessary condition for the existence of a limit of restoration
of $N=4$ supersymmetry is the vanishing of $B_2$.
In this limit, $B_4$ receives a non-zero
contribution only from one $N=4$ sector of the orbifold.
When there is a restoration of $N=8$, also 
the massless contribution to $B_4$ must vanish. 
As is clear from our formulae, however,
this implies the full vanishing of this helicity supertrace.

\noindent

\vskip 0.3cm
\setcounter{section}{5}
\setcounter{equation}{0}
\section*{\normalsize{\bf 5. Perturbative and non-perturbative dualities}}

The knowledge of the partition function allows us to analyse
many properties of the string constructions.
In this section we consider perturbative and non-perturbative
string--string dualities.

\subsection*{\normalsize{\sl 5.1. Mirror symmetry  
from the partition function}}

In Appendix B.2  we illustrate how to pass 
from type IIA to type IIB in the
framework of the fermionic construction. Here we want to
show how mirror symmetry, namely the statement that the type IIA
string compactified on the Calabi--Yau manifold $M$ is
equivalent to the type IIB string compactified on the mirror 
manifold $\tilde{M}$ \footnote{See for instance \cite{kt}.},
can be easily read off at the orbifold points we are considering.
In order to see this,
we start by going to the fermionic point of the moduli space of the 
orbifolds.
At such a point, as we saw,
the operation of passing from a space 
$M$ to the mirror $\tilde{M}$ is implemented by a change in
the modular coefficient $C_{(b_1|b_2)}$, which is responsible
for the sign of $B_2=N_V-N_H=- \chi/2$.
On the other hand, passing from IIA to IIB requires the changes 
quoted in Appendix B.2, which involve also a change of sign of $B_2$. 
When combined, the two operations of exchanging IIA with IIB
and $M$ with $\tilde{M}$ leave $B_2$ invariant and exchange
$T^i$ with $U^i$ in $B_4$.
However, at the level of the partition function,
such an exchange simply amounts to a different choice of 
the vectors $w$, which specify, for each plane, the lattice shifts.
The initial choice of $w$ is arbitrary.
In particular, we can choose $w$ in such a way that the 
quantities (\ref{O})--(\ref{DD}), which encode all the 
non-trivial moduli dependence of the models, are
invariant under the exchange of $T$ with $U$: all the other
choices are related to that by $SL(2,Z)$ transformations in 
$T$ and/or $U$. We therefore see that, modulo
$SL(2,Z)$ transformations,
$B_2$ and $B_4$ are invariant under mirror symmetry.
To conclude that this is a perturbative symmetry of the theory,
it is then sufficient to observe that the pair $(B_2,B_4)$ is in a
one-to-one
correspondence with the partition function. This means that this pair
uniquely determines the model,
encoding all the perturbative physics at any order 
of perturbation.

\subsection*{\normalsize{\sl 5.2. String--string U-dualities }}

The non-perturbative dualities we consider here are string--string U-dualities,
which relate the type IIA orbifolds to heterotic or type II duals.
The type II duals are constructed as asymmetric orbifolds in which
the $N=2$ supersymmetry is realized only among left-movers.
As for the heterotic constructions, also in these type II orbifolds
the dilaton--axion field belongs to the vector manifold
(see Appendix B.3); it
is exchanged by U-duality with one of the moduli of the
vector manifold of the type IIA duals. 
This kind of duality is therefore much similar to the
duality between type IIA and heterotic strings. In the case of the 
heterotic/type IIA duality, 
a necessary condition for the identification of the moduli
is the compactification of the type IIA
string on a $K3$ fibration \cite{al}. 
When the conformal field theory can be explicitly 
solved, as in our $Z_2 \times Z_2$ orbifolds, this requirement 
translates into the property 
of spontaneous breaking of the $N=4$ supersymmetry,
in the sense we described above\footnote{
The connection relies on the fact that in the limit of large
volume of the compact space, any $K3$ fibration looks locally
like ${\bf C} \times K3$. This means that locally, an observer sitting
on a point of the base  sees 16 supercharges,
as in the $T^2 \times K3$ compactification, instead of 8.
The extra 8 supercharges are projected out by global, not local,
projection.
In the case of $Z_2 \times Z_2$ orbifolds, the  $K3$ 
is described by the orbifold limit $T^4/Z_2$: 
the $Z_2 \times Z_2$ orbifolds which correspond to $K3$ fibrations are those
for which, in some corner of
the moduli $T$, $U$, which always corresponds to the decompactification
of some dimensions, one of the two $Z_2$ projections can be made
to effectively vanish, thereby recovering a $T^2 \times T^4/Z_2$, $N=4$
orbifold.
In all such models, the $N=2$ and $N=4$ phases are continuously
related by a change of size, or shape, of the compact space.
The $N=4$ phase is reached when at least one dimension is decompactified.}.

Therefore the models that are orbifold limits of $K3$ fibrations are the 
following:
\be
\begin{array}{ccl}
(8,8) & ~~~~~ & (O,F,F) \\
      && \\
(4,4) & ~~~~~ & (D,FD,F^*) \\
      &&\\
(2,2) & ~~~~~ & (DD,FD,F^*) \\
      &&\\
(0,0) & ~~~~~ & (F,F,F) \\
      & ~~~~~ & (FD,FD,F^*) 
\end{array}
\ee
(here $F^*$ stands for $F$ as well as for $FD$).
The heterotic duals of the first three models were considered
in \cite{fhsv,gkp,hmn=2}.
The model $(0,0)$ is special, possessing also a spontaneously
broken $N=8$ supersymmetry. The detailed study of this last model
has been considered in \cite{gkp2}.

In the case of type IIA/type II asymmetric orbifold duality,
on which we will concentrate in the following,
the recipe is not the compactification on a $K3$ fibration
(or some of its orbifold limits). However, 
the type IIA/heterotic $U$-duality 
can be tested by looking at the renormalization of certain terms in the
effective action. In \cite{gkp},
a particular linear combination of $R^2$ and 
$F_{\mu \nu}F^{\mu \nu}$ terms,
smooth and analytic in the full space of moduli $T$ and $U$,
was shown to be appropriate for a 
comparison of type IIA and heterotic constructions.
The same combination of gravitational and gauge field-strengths
can be used here as a guideline in the search of
type IIA/type II asymmetric dual pairs.
Actually, since in the type II constructions all the gauge bosons are
Ramond--Ramond states, the above amplitude turns out to coincide with
the $R^2$ term alone. As it happens for the type IIA,
also in the type II asymmetric orbifold constructions the genus zero
contribution to this term vanishes. 
In the type II effective action there is therefore no bare coupling constant, 
and the dilaton contribution
is only non-perturbative and exponentially suppressed.
Such a behaviour is reproduced on type IIA by the moduli 
$T^i$ of the planes denoted, in our convention, by $F$ or $FD$.
Their contribution to the renormalization of the $R^2$ term
is therefore
\be
\log \Im T \left\vert \th_4 \left( T \right) \right\vert^4
\longrightarrow \log \Im T~~\left( \sim \, 0 \right)~+~
O \left( {\rm e}^{-iT} \right)~~~~~~~\left( \left\vert T \right\vert 
\to \infty \right). 
\label{thetat}
\ee
The mild logarithmic divergence is an infrared artefact 
and can be lifted by switching on an  
appropriate cut-off (these planes behave in fact like the planes shifted
by the projection that spontaneously breaks the supersymmetry
in the models of Refs. \cite{6auth,gkp,gkp2}.)

Once the plane whose K\"{a}hler class modulus $T$
is mapped into the dilaton field of the asymmetric orbifold
has been identified, the contribution of the moduli 
of the remaining planes has to match
the contribution of the perturbative vector multiplet
moduli in the type II asymmetric orbifold.
It turns out that
the models that possess such an asymmetric dual construction are
\be
\begin{array}{ccl}
(16,16) & ~~~~~ & (F,O,O) \\
      && \\
(8,8) & ~~~~~ & (F^*,D,D) \\
      &&\\
(4,4) & ~~~~~ & (F^*,DD,DD) \\
      &&\\
(0,0) & ~~~~~ & (F,F,F) \\
      & ~~~~~ & (F^*,FD,FD) \, . 
\end{array}
\label{asdual}
\ee
Among these, a special role is still played by the model $(0,0)$,
which therefore possesses both a heterotic and a type II asymmetric
dual. In this model, 
one of the moduli $T$ is mapped in the dilaton of the
asymmetric construction and in the inverse of the dilaton of the heterotic dual
(in the limit $T \to 0$, (\ref{thetat}) shows up
a linear behaviour in $\tilde{T}\equiv -1/T$, which matches the
tree level ${1 \over g^2} \sim \Im S$ correction on the heterotic side
\cite{gkp2}).
Since this modulus plays the role of a Higgs field for the
spontaneous breaking of some of the supersymmetries in the type IIA
orbifold, we learn through the duality map 
that there is, on the heterotic side, a
non-perturbative spontaneous breaking of an $N=8$ supersymmetry
\cite{gkp2}.
Similar arguments can be applied to the first three models
of (\ref{asdual}), namely the constructions with $(N_V,N_H)$ $=$
$(16,16)$, $(8,8)$ and $(4,4)$,
which we analyse here in detail. It is possible to show 
that also in such models there is a non-perturbative super-Higgs phenomenon.
The dependence on the dilaton $S^{\rm II}$, in these cases,
can be obtained by looking at the asymmetric duals.
The dual of $S^{\rm II}$ is in fact a perturbative modulus
belonging to a hypermultiplet, whose dependence is explicit
in the asymmetric constructions; it is not difficult to
identify the latter with one of the super-Higgs fields
responsible for the spontaneous breaking of some of the supersymmetries. 
Through the duality between symmetric and asymmetric
orbifolds, we therefore learn that, 
in the strong coupling limit, these type IIA 
orbifolds have an approximate restoration of a $N=4$ supersymmetry.

In order to see the above issues in detail,
we start by discussing the type IIA orbifolds.
In these specific cases, Eq. (\ref{ztu}) reads:
\ba
Z_{\rm II}^{(1,1)} & = &
{1 \over \Im \tau \vert \eta \vert^{24} } {1 \over 4}
\sum_{H^{\rm o},G^{\rm o}} \sum_{H^{\rm f},G^{\rm f}}
\Gamma_{6,6}^{N_V} \ar{H^{\rm o},H^{\rm f}}{G^{\rm o},G^{\rm f}} \nonumber \\
&& \times {1 \over 2} ~\sum_{a,b} (-)^{a+b+ab}~ \vartheta \ar{a}{b}
\vartheta \ar{a+H^{\rm o}}{b+G^{\rm o}}
\vartheta \ar{a+H^{\rm f}}{b+G^{\rm f}}
\vartheta \ar{a-H^{\rm o}-H^{\rm f}}{b-G^{\rm o}-G^{\rm f}} \nonumber \\
&& \times {1 \over 2} \sum_{\bar{a},\bar{b}} 
(-)^{\bar{a}+\bar{b}+\bar{a}\bar{b}} \bar{\vartheta} \ar{\bar{a}}{\bar{b}}
\bar{\vartheta} \ar{\bar{a}+H^{\rm o}}{\bar{b}+G^{\rm o}}
\bar{\vartheta} \ar{\bar{a}+H^{\rm f}}{\bar{b}+G^{\rm f}}
\bar{\vartheta} \ar{\bar{a}-H^{\rm o}-H^{\rm f}}{\bar{b}-G^{\rm o}-G^{\rm f}}~.
\label{part}
\ea
The characters $\Gamma_{6,6}^{N_V} \ar{H^{\rm o},H^{\rm f}}
{G^{\rm o},G^{\rm f}}
\equiv |\eta|^{12} Z_{6,6}$, are given by: 
\ba
\Gamma_{6,6}^{16} \ar{H^{\rm o},H^{\rm f}}{G^{\rm o},G^{\rm f}} & = &
\Gamma_{2,2}^{(1)} \ar{H^{\rm o}~ \vert~H^{\rm f}}{G^{\rm o}~ 
\vert~G^{\rm f}}  
\Gamma_{2,2}^{(2)} \ar{H^{\rm f}~\vert~0}{G^{\rm f}~\vert~ 0}  
\Gamma_{2,2}^{(3)} \ar{H^{\rm o}+H^{\rm f}~ \vert~0}
{G^{\rm o}+G^{\rm f}~ \vert~0}~, \label{bos1} \\
&& \nn \\
\Gamma_{6,6}^{8} \ar{H^{\rm o},H^{\rm f}}{G^{\rm o},G^{\rm f}}& = &
{1 \over 2}\, \sum_{H^{D_1},G^{D_1}}~
\Gamma_{2,2}^{(1)} \ar{H^{\rm o}~ \vert~H^{\rm f}~;~0}
{G^{\rm o}~ \vert~G^{\rm f}~;~0}  
\Gamma_{2,2}^{(2)} \ar{H^{\rm f}~\vert~H^{D_1}}{G^{\rm f}~\vert~ G^{D_1}}  
\nonumber \\
&& \times~ \Gamma_{2,2}^{(3)} \ar{H^{\rm o}+H^{\rm f}~ \vert~H^{D_1}}
{G^{\rm o}+G^{\rm f}~ \vert~G^{D_1}}~, \label{bos2} \\
&& \nn \\
{\rm and}~~~~~~~~~~~~~~~~ && \nn \\
&& \nn \\
\Gamma_{6,6}^{4} \ar{H^{\rm o},H^{\rm f}}{G^{\rm o},G^{\rm f}} & = &
{1 \over 4}\, \sum_{H^{D_1},G^{D_1}}~\sum_{H^{D_2},G^{D_2}}~
\Gamma_{2,2}^{(1)} \ar{H^{\rm o}~ \vert~H^{\rm f}~;~0}
{G^{\rm o}~ \vert~G^{\rm f}~;~0}\nonumber \\  
&& \times~\Gamma_{2,2}^{(2)} \ar{H^{\rm f}~\vert~0~;~H^{D_1},~H^{D_2}}
{G^{\rm f}~\vert~ 0~;~G^{D_1},~G^{D_2}}  
\Gamma_{2,2}^{(3)} \ar{H^{\rm o}+H^{\rm f}~ \vert~0~;~H^{D_1},~H^{D_2}}
{G^{\rm o}+G^{\rm f}~ \vert~0~;~G^{D_1},~G^{D_2}}\, , \label{bos3}  
\ea
where $(H^{\rm o},G^{\rm o})$ 
refer to the boundary conditions introduced by the
projection $Z_2(b_2)$ (see Section 2)
and $(H^{\rm f},G^{\rm f})$ refer 
to the projection $Z_2(b_1)$, which acts freely, as a
rotation in the complex planes 2, 3
and a translation, ${\rm e}^{i \pi m_2 G^{\rm f}} $, in the lattice of
the first complex plane. In (\ref{bos2}), (\ref{bos3})
we used the generalized characters that include the action 
of the $D$-projections as they were defined in the Eq. (\ref{gts}) and
the following.
The corresponding helicity supertraces $B_4$ 
are\footnote{ We recall that the prime summation over 
$(h,g)$ stands for $(h,g)=\{(0,1),(1,0),(1,1) \}$.}:
\ba
B_4^{N_V=16} & = & 6 \sump \Gamma_{2,2}^{(1)} \ar{0|h}{0|g}+
18 \sum_{i=2,3} \Gamma_{2,2}^{(i)}~, \label{b41} \\
&& \nn \\
B_4^{N_V=8} & = & 3 \sump \left( \Gamma_{2,2}^{(1)} \ar{0|h,0}{0|g,0}
+ \Gamma_{2,2}^{(1)} \ar{0|h,h}{0|g,g} \right) \nonumber \\
&& + 9 \Gamma_{2,2}^{(2)} + 
3 \sump \Gamma_{2,2}^{(2)} \ar{0|h}{0|g} + 
9 \Gamma_{2,2}^{(2)} + 
3 \sump \Gamma_{2,2}^{(3)} \ar{0|h}{0|g}~, \label{b42} \\ 
&& \nn \\
B_4^{N_V=4} & = & 3 \sump \left( \Gamma_{2,2}^{(1)} \ar{0|h,0}{0|g,0}
+ \Gamma_{2,2}^{(1)} \ar{0|h,h}{0|g,g} \right) \nonumber \\
&& + {9 \over 2} \Gamma_{2,2}^{(2)}+ 
{3 \over 2} \sump \left(\Gamma_{2,2}^{(2)} \ar{0|h,0}{0|g,0}+  
\Gamma_{2,2}^{(2)} \ar{0|0,h}{0|0,g}+\Gamma_{2,2}^{(2)} \ar{0|h,h}{0|g,g}
\right) \nonumber \\
&& + {9 \over 2} \Gamma_{2,2}^{(3)}+ 
{3 \over 2} \sump \left(\Gamma_{2,2}^{(3)} \ar{0|h,0}{0|g,0}+  
\Gamma_{2,2}^{(3)} \ar{0|0,h}{0|0,g}+\Gamma_{2,2}^{(3)} \ar{0|h,h}{0|g,g}
\right)~. \label{b43} 
\ea
In order to obtain the gravitational corrections,
we proceed as in \cite{gkp,gkp2}: the four derivative gravitational 
corrections we will consider are precisely 
those that were analysed in the framework
of $N=4$ ground states of Ref. \cite{6auth} and the $N=2$
ground states of Refs. \cite{gkp,gkp2}.
There is no tree-level contribution to these operators, and the $R^2$ 
correction is related
to the insertion of the two-dimensional operator 
$2 \lambda^2 \bar \l^2$ in the one-loop partition function.
In the models at hand, since supersymmetry is
realized symmetrically and $N_V=N_H$, the
contribution of the $N=2$ sector to $B_4$ vanishes, and 
$\left\langle 2\l^2 \bar \l^2 \right\rangle$ is identified
with $B_4 / 3$. The massless contributions of the latter
give rise to an infrared logarithmic behaviour $2b_{\rm
II}\log[M^{(\rm IIA)\,2}  /\mu^{(\rm IIA)\, 2}]$ \cite{infra,delgrav}, where
$M^{(\rm IIA)}\equiv\frac{1}{\sqrt{\alpha'_{\rm IIA}}}$ is
the type IIA string scale and $\mu^{(\rm IIA)}$ is the type IIA
infrared cut-off.
Besides this running, the one-loop correction contains the
thresholds $\Delta_{\rm IIA}$, which account for the infinite tower of
string modes.

The one-loop corrections of the $R^2$-term are then related to the
infrared-regularized genus-one
integral of $B_4 / 3$. In the type IIA
string, these $R^2$ corrections depend on the K\"ahler moduli
(spanning the vector manifold), and are independent of the
complex-structure moduli (spanning the scalar manifold):
\be
\partial_{T^i}\Delta_{\rm IIA}=\frac{1}{3}\ifd
\partial_{T^i} B_4
\ , \ \
\partial_{U^i}\Delta_{\rm IIA}=0\, .
\label{IIAthr}
\ee
In  the  type IIB string, the roles of $T^i$ and $U^i$ are
interchanged. We obtain the following one-loop correction
to the coupling constant:
\be
{ 16 \, \pi^2 \over g^2_{\rm grav} \left( \mu^{(\rm IIA)} \right)}=
- 2\log \Im T^1 
\left\vert  \th_4 \left( T^1 \right) \right\vert^4 ~+~
\Delta^{N_V} \left( T^2,T^3 \right)
~+ \left( 6 +{N_V \over 2} \right) \log {M^{(\rm IIA)}  \over 
\mu^{(\rm IIA)} }~,
\label{IIthr}
\ee
where the various ``thresholds'' $\Delta^{N_V} \left( T^2,T^3 \right)$ read:
\ba
\Delta^{16} \left( T^2,T^3 \right)& = &
-6 \log \Im T^2 \left\vert  \eta \left(  T^2    \right) \right\vert^4
-6 \log \Im T^3 \left\vert  \eta \left(  T^3    \right) \right\vert^4~,
\label{dIIA1}\\
&& \nn \\
\Delta^{8} \left( T^2,T^3 \right) &=&
-3  \log \Im T^2 
\left\vert  \eta \left(  T^2    \right) \right\vert^4
- \log \Im T^2 
\left\vert  \th_4 \left(  T^2    \right) \right\vert^4\nn \\
&& -3  \log \Im T^3 
\left\vert  \eta \left(  T^3    \right) \right\vert^4
- \log \Im T^3 
\left\vert  \th_4 \left(  T^3    \right) \right\vert^4~, \\
&& \nn \\
\Delta^{4} \left( T^2,T^3 \right) &=&
-{3 \over 2} \log \Im T^2 
\left\vert  \eta \left(  T^2    \right) \right\vert^4
-{3 \over 2} \log \Im T^2 
\left\vert  \th_4 \left(  T^2    \right) \right\vert^4\nn \\
&& -{3 \over 2} \log \Im T^3 
\left\vert  \eta \left(  T^3    \right) \right\vert^4
-{3 \over 2} \log \Im T^3 
\left\vert  \th_4 \left(  T^3    \right) \right\vert^4~.
\ea
In the last model, $N_V=N_H=4$, if the semi-freely acting
projection on the third complex plane is a product of $\tilde{D}$-
instead of $D$-operations we obtain:
\ba
\Delta^{4} \left( T^2,T^3 \right) &=&
-{3 \over 2} \log \Im T^2 
\left\vert  \eta \left(  T^2    \right) \right\vert^4
-{3 \over 2} \log \Im T^2 
\left\vert  \th_4 \left(  T^2    \right) \right\vert^4\nn \\
&& -3 \log \Im T^3 
\left\vert  \eta \left(  T^3    \right) \right\vert^4~.
\label{dIIA3}
\ea
Notice that, except for the planes 2 and 3 of model $N_V=16$, 
the shifts on the $\Gamma^{(i)}_{2,2}$ lattices break
the $SL(2,Z)_{T^i}$ duality groups. As in \cite{gkp,gkp2}, the actual subgroup 
left unbroken depends on the kind of shifts performed
(see Refs. \cite{6auth, kkprn,solving}).
All the above corrections diverge linearly, in both the
large and small $T^2$ and $T^3$ limits.
On the other hand, the contribution of $T^1$ diverges
only logarithmically in the large-$\Im T^1$ limit, and
linearly in the inverse modulus $\tilde{T}=-1/T^1$, for small $T^1$.
As we previously discussed,
the logarithmic divergence is an infrared artefact 
and can be removed by switching on an appropriate cut-off.

\subsection*{\normalsize{\sl 5.3. The type II asymmetric duals}}

We now discuss the type II asymmetric dual orbifolds.
The model $N_V=N_H=16$ is constructed starting from
the $N=8$ IIA superstring compactified on $T^6$ and
applying two projections: $Z_2^{( F)}$ and $Z_2^{(\rm o)}$.
$Z_2^{( F)}$ acts freely, as $(-)^{F_{\rm R}}$ together
with a translation on $T^6$, and 
projects out all the left-moving supersymmetries. 
$Z_2^{(\rm o)}$, instead, acts as a rotation that reduces
symmetrically the number of  supersymmetries by $1/2$.
The properties of the $N=4$ models obtained by applying only 
$Z_2^{ (F)}$ were already analysed in \cite{6auth}.
The orbifold obtained by the further application of $Z_2^{(\rm o)}$ 
has an $N=2$ supersymmetry, which is realized only among left-movers.
The partition function of the model reads
\ba
Z_{\rm II}^{(2,0)} & =  & {1 \over \Im \tau |\eta|^{24} }
{1 \over 4} \sum_{H^{\rm F},G^{\rm F}} \sum_{H^{\rm o},G^{\rm o}}
\Gamma_{6,6} \ar{H^{\rm F}, H^{\rm o}}{G^{\rm F}, G^{\rm o}} \nonumber \\
&& \times {1 \over 2}
\sum_{a,b}(-)^{a+b+ab} 
\vartheta^2 \ar{a}{b}
\vartheta \ar{a+H^{\rm o}}{b+G^{\rm o}}\vartheta \ar{a-H^{\rm o}}{b-G^{\rm o}}
 \nonumber \\
&& \times {1 \over 2} 
\sum_{\bar{a},\bar{b}}(-)^{\bar{a}+\bar{b}+\bar{a}\bar{b}} 
(-)^{\bar{a}G^{\rm F}+\bar{b}H^{\rm F}+H^{\rm F}G^{\rm F}}
\bar{\vartheta}^2 \ar{\bar{a}}{\bar{b}} 
\bar{\vartheta} \ar{\bar{a}+H^{\rm o}}{\bar{b}+G^{\rm o}}
\bar{\vartheta} \ar{\bar{a}-H^{\rm o}}{\bar{b}-G^{\rm o}}~, 
\label{z2as}
\ea
where now
\be
\Gamma_{6,6} \ar{H^{\rm F}, H^{\rm o}}{G^{\rm F}, G^{\rm o}} =
\Gamma_{2,2}^{(1)} \ar{H^{\rm o}~|~H^{\rm F}}{G^{\rm o}~|~G^{\rm F}}
\Gamma_{2,2}^{(2)} \ar{H^{\rm o}~|~0}{G^{\rm o}~|~0}
\Gamma_{2,2}^{(3)} \ar{0~|~0}{0~|~0}~.
\label{z20}
\ee
Notice that neither $Z_2^{(\rm F)}$ nor $Z_2^{(\rm o)}$
act on the third plane.
The massless spectrum can be easily analysed by computing
the helicity supertraces $B_2$ and $B_4$.
$B_2$ turns out to be zero (for the details of this computation, 
see the Appendix F). This tells us that $N_V=N_H$.
The supertrace $B_4$ is (cf. \cite{gkp2}):
\ba
B_4 & = & { 3 \over 16}~ {1 \over \bar{\eta}^{12}}~ \sum_{\bar{a},\bar{b}} 
\sum_{(H^{\rm F},G^{\rm F})}(-)^{\bar{a}+\bar{b}+\bar{a}\bar{b}} 
(-)^{\bar{a} G^{\rm F}+\bar{b} H^{\rm F} +H^{\rm F} G^{\rm F}}
\bar{\vartheta}^4 \ar{\bar{a}}{\bar{b}} 
\Gamma_{6,6} \ar{H^{\rm F},0}{G^{\rm F},0}
\nonumber \\
&& +~ 36 \, \Gamma_{2,2}^{(3)} \ar{0~|~0}{0~|~0}
\label{B420}
\ea 
The massless contribution, which coincides with the supergravity result, is
\be
B_4\vert_{\rm massless}=
18+{7N_V-N_H \over 4}=42~,
\ee
from which we derive $N_V=N_H=16$.
The massless spectrum therefore contains, besides the supergravity multiplet,
3+16 vector multiplets and 4+16 hypermultiplets: it is therefore the same
as that of the type IIA model $N_V=16$. However, here
the dilaton belongs to a vector multiplet.
This is a general property of all the $N=2$ string compactifications
in which all the supersymmetries are realized
either between only left or between only right movers,
such as the heterotic strings or  type II asymmetric orbifolds
as the ones we consider. The reason is that
the dilaton, in such cases, is uncharged under the
$SU(2)$ operators that rotate the supercharges of the $N=2$ 
supergravity\footnote{ One can construct the vertex operators, which represent 
the states of the string theory, and then construct 
explicitly the generators of supersymmetry in this representation
(see Appendix B.3).
One then finds that the spinor vertex operator, which corresponds to the
two transverse space-time coordinates
${\rm e}^{-{i \over 2}H_0}$, is common to both
the supercharges (in type II symmetric orbifolds, we have instead
$Q \sim {\rm e}^{-{i \over 2}H_0}$,  
$\bar{Q} \sim {\rm e}^{-{i \over 2}\bar{H}_0}$). 
As a consequence, the generators of the $SU(2)$ 
symmetry of the $N=2$ do not act on the space-time degrees of freedom,
and therefore leave invariant
the dilaton, whose vertex operator actually contains
only space-time degrees of freedom.}.

The $N_V=N_H=8$ orbifold is constructed by modding out the previous model
with a further $Z_2^{(D)}$ projection, which acts semi-freely.
The partition function of this orbifold is given as in (\ref{z2as}),
but with (\ref{z20}) replaced by
\be
\Gamma_{6,6} \ar{H^{\rm F},H^{\rm o}}{G^{\rm F},G^{\rm o}}=
{1 \over 2} \sum_{(H^{D_1},G^{D_1})}
\Gamma^{(1)}_{2,2} 
\ar{H^{\rm o}~\vert~H^{\rm F}~;~0}{G^{\rm o}~\vert~G^{\rm F}~;~0}
\Gamma^{(2)}_{2,2} 
\ar{H^{\rm o}~\vert~0~;~H^{D_1}}{G^{\rm o}~\vert~0~;~G^{D_1}}
\Gamma^{(3)}_{2,2} 
\ar{0~\vert~0~;~H^{D_1}}{0~\vert~0~;~G^{D_1}}~.
\ee 

The helicity supertrace $B_4$ is now given by an expression similar
to (\ref{B420}), but the second term, instead of being
$36 \Gamma_{2,2}^{(3)} \ar{0|0}{0|0}$, is now 
\be
12 \sump \left({1 \over 2} \Gamma_{2,2}^{(3)}\ar{0~\vert~0}{0~\vert~0}+  
{1 \over 2} \Gamma_{2,2}^{(3)}\ar{0~\vert~h}{0~\vert~g} \right)~,
\ee
which gives 
\be
B_4 \vert_{\rm massless}=30
\ee
consistently with $N_V=8$ ($B_2=N_V-N_H$ is zero in all these models).

Finally, the model $N_V=N_H=4$ is obtained by applying
two $Z_2^{(D)}$ projections, which act on the compact space, producing the
following character:
\ba
\Gamma_{6,6} \ar{H^{\rm F},H^{\rm o}}{G^{\rm F},G^{\rm o}}&=&
{1 \over 4} \sum_{(H^{D_1},G^{D_1})}\sum_{(H^{D_2},G^{D_2})}
\Gamma^{(1)}_{2,2} 
\ar{H^{\rm o}~\vert~H^{\rm F}~;H^{D_1}}
{G^{\rm o}~\vert~G^{\rm F}~;~G^{D_1}}\nn \\
&& \times \,
\Gamma^{(2)}_{2,2} 
\ar{H^{\rm o}~\vert~0~;~H^{D_2}}{G^{\rm o}~\vert~0~;~G^{D_2}}
\Gamma^{(3)}_{2,2} 
\ar{0~\vert~0~;~H^{D_1},H^{D_2} }
{0~\vert~0~;~G^{D_1},G^{D_2} }~.
\ea 
In this case, the second term in $B_4$ is
\be
12 \sump \left({1 \over 4} \Gamma_{2,2}^{(3)}\ar{0~\vert~0,0}{0~\vert~0,0}+  
{1 \over 4} \Gamma_{2,2}^{(3)}\ar{0~\vert~h,0}{0~\vert~g,0}+
{1 \over 4} \Gamma_{2,2}^{(3)}\ar{0~\vert~0,h}{0~\vert~0,g}+
{1 \over 4} \Gamma_{2,2}^{(3)}\ar{0~\vert~h,h}{0~\vert~g,g}
\right)~,
\ee
which gives $B_4 \vert_{\rm massless}=24$ and $N_V=4$.

Owing to the free action of $Z_2^{ F_{\rm R}}$,
all these models possess a spontaneously broken $N=4=(2,2)$ supersymmetry,
restored in the limit in which
the projection $Z_2^{ F_{\rm R}}$ becomes irrelevant: this
phenomenon takes place
for special values of the moduli belonging to hypermultiplets.
We remark that in all these models $B_2 \equiv 0$.
As a consequence there are no points in the moduli space in which
there appear ``$N=2$'', $\Delta N_V \neq \Delta N_H$ singularities.

As in the case of the type IIA orbifolds, also in the asymmetric duals
the $R^2$
gravitational corrections receive a contribution only from one loop,
and are obtained  by
the insertion of the operator $\lambda^2_{\rm L} \lambda^2_{\rm R}$
(see \cite{gkp2}).
The only non-zero contribution is provided by the sectors with
$(H^{\rm o},G^{\rm o}) \neq (0,0)$. 
We obtain:
\be
{16 \, \pi ^2 \over g^2_{\rm grav} \left( \mu^{(\rm As)} \right)} =
\left(4+{N_V \over 2} \right) 
\log {M^{(\rm As )} \over \mu^{(\rm As )}}+
\Delta^{N_V} \left(T^{\rm As},U^{\rm As} \right)~,
\ee
where we introduced the type II asymmetric mass scale and
infrared cut-off, $M^{(\rm As )}$ and $\mu^{(\rm As )}$ respectively,
and
\ba
\Delta^{16} \left(T^{\rm As},U^{\rm As} \right) &=&
-6 \log \Im T^{\rm As} \left\vert \eta \left( T^{\rm As} \right) 
\right\vert^4\, -
6 \log \Im U^{\rm As} \left\vert \eta \left( U^{\rm As} \right) 
\right\vert^4\,; \label{dAs1}\\
&& \nn \\
\Delta^{8} \left(T^{\rm As},U^{\rm As} \right) \, & = &
- 3  \log \Im T^{\rm As}\left\vert \eta \left( T^{\rm As} \right) 
\right\vert^4\,
- 3  \log \Im U^{\rm As}\left\vert \eta \left( U^{\rm As} \right) 
\right\vert^4 \nn \\
&& - \log \Im T^{\rm As}\left\vert \th_i 
\left( T^{\rm As} \right) \right\vert^4\,
- \log \Im U^{\rm As}\left\vert \th_j \left( U^{\rm As} \right) 
\right\vert^4 \, ; \\ 
&& \nn \\
\Delta^{4} \left(T^{\rm As},U^{\rm As} \right) \, & = &
- 3  \log \Im T^{\rm As}\left\vert \eta \left( T^{\rm As} \right) 
\right\vert^4\,
- {3 \over 2} \log \Im U^{\rm As}\left\vert \eta \left( U^{\rm As} \right) 
\right\vert^4 \nn \\
&& - {3 \over 2} \log \Im T^{\rm As}\left\vert \th_4 
\left( T^{\rm As} \right) \right\vert^4\, . 
\label{dAs3}
\ea

\subsection*{\normalsize{\sl 5.4. Comparison of symmetric and asymmetric
orbifolds }}

We now come to the comparison of the
type IIA symmetric and the type II asymmetric orbifolds.
As is clear from expressions (\ref{dIIA1})--(\ref{dIIA3})
and the analogous (\ref{dAs1})--(\ref{dAs3}) for the 
asymmetric orbifolds, for any type IIA symmetric orbifold it is possible
to choose actions of the $D$-projections 
that can be reproduced in the type II asymmetric orbifolds,
leading to the same corrections $\Delta^{N_V}$:
$\Delta^{N_V}(T^2,T^3)=\Delta^{N_V}(T^{\rm As},U^{\rm As})$. 
A comparison of the corrections with the $R^2$ term
therefore leads to the following identification of 
the moduli in the vector manifold:
$T^2 =T^{\rm As}$, $T^3 = U^{\rm As}$.
These moduli are perturbative in both the constructions.
On the other hand,
the contribution of the modulus $T^1$ in (\ref{IIthr}),
in the limit $T^1 \to \infty$,
diverges only logarithmically. This is consistent with the vanishing
of the perturbative, genus-zero ($O(S^{\rm As})$) contribution
to this term in the type II asymmetric orbifolds.      
We are therefore led to 
identify this modulus with the
dilaton--axion field $\tau^{\rm As}_S \equiv 4 \pi S^{\rm As}$ 
of the asymmetric orbifolds \cite{6auth,gkp2}.
The logarithmic behaviour is then interpreted as a non-perturbative effect.

We remark that the identification of the perturbative moduli
is possible only for special choices of the  translations
introduced by the $D$-projections. 
The reason is that it is not always possible to reconstruct
the properties of the K\"{a}hler class moduli of a product
of two tori, whose lattice of momenta and windings is
shifted by the action of the $D$-projections, with
the K\"{a}hler class and complex structure moduli of
a single torus with shifted lattice:
different translations correspond to different cuts
in the moduli space of the model,
and not all the cuts correspond to dual constructions.
However, we learned that a correct cut exists, and the identification
of $(T^2,T^3)$ with $(T^{\rm As},U^{\rm As})$
provides a test of the duality. 
The corrections computed in the type IIA symmetric orbifolds
therefore provide the full, perturbative and non-perturbative, corrections
for the asymmetric orbifolds. 
These are given by the expression (\ref{IIthr}), 
in which \\
\vspace{.1cm}
\noindent
\romannumeral 1)
the moduli
$T^1,T^2$ are replaced by $T^{\rm As},U^{\rm As}$,
\\
\vspace{.1cm}
\noindent
\romannumeral 2) the modulus $T^1$ is replaced by the dilaton 
$4 \pi S^{\rm As}$,
\\
\vspace{.1cm}
\noindent
\romannumeral 3) the type IIA string mass and cut-off have to be
replaced by those of the type II asymmetric theory. Indeed,
we are using a regularization scheme for which the ratio of the mass and the 
cut-off is duality-independent, and can be expressed in terms of
the ratio of the Plank mass and a physical cut-off $\mu$:
\be
{M^{(\rm IIA)} \over \mu^{(\rm IIA)}}=
{M^{(\rm As)} \over \mu^{(\rm As)}}=
{M_{\rm Planck} \over \mu}~.
\label{mmu}
\ee
This duality also provides, on the other hand, a window on the
non-perturbative physics of the type IIA orbifolds.
As we already pointed out, in the above type IIA orbifolds 
supersymmetry is broken in a ``rigid'' way; namely, it is not possible
to restore some of the broken supersymmetries by taking
some special limit in the moduli of the compact space.
This has to be contrasted with the situation of the type II
asymmetric dual orbifolds, in which instead the
asymmetric projection $Z_2^{\rm F_{\rm R}}$ acts freely, and it can be
made to effectively vanish by taking an appropriate limit in the
moduli of the first plane. These are
hypermultiplet moduli. 
In these orbifolds, the first and the second plane provide
four such moduli, that correspond to the three hypermultiplet
moduli $U^1$, $U^2$, $U^3$ and the
dilaton $S^{\rm II}$ of the type IIA orbifolds.
Three moduli are therefore perturbative in both the theories, and
the mismatch in the perturbative mechanism of spontaneous 
breaking of supersymmetry forces us to identify the modulus
of the asymmetric orbifolds, which plays the role of super-Higgs field
as the dual of the type IIA dilaton.
{}From the point of view of type IIA,
the mechanism of spontaneous breaking of supersymmetry is then
entirely non-perturbative.

\noindent

\vskip 0.3cm
\setcounter{section}{6}
\setcounter{equation}{0}
\section*{\normalsize{\bf 6. Conclusions}}

In this work we  constructed and classified all the four-dimensional 
$N=2$, $Z_2 \times Z_2$ 
symmetric orbifolds of the type IIA/B superstring.
After having constructed the models at the fermionic point,
according to the rules of the ``fermionic construction'', we
established the equivalence, for this class of 
orbifolds, of world-sheet fermions and bosons, and we derived the partition
function of each model at a generic point in the space of the moduli of
the three tori of $T^6=T^2 \times T^2 \times T^2$.
Through an analysis of some helicity supertraces, 
easily computable from the partition function,
we investigated the appearance of stringy Higgs and super-Higgs
phenomena, which served as a guideline in the search
for heterotic and/or type II duals.
We devoted particular attention to the study of the latter,
for which we provided an analysis analogous to that of the
symmetric orbifolds. These pairs are related by a map that exchanges
the dilaton--axion field of the type II asymmetric construction
for a perturbative modulus of the type IIA dual, associated to
a vector multiplet. Conversely, the type IIA dilaton is mapped
into a perturbative modulus of the type II asymmetric dual,
associated to a hypermultiplet.
Through the comparison of the corrections to the $R^2$ term,
we then provided a test of this duality, 
obtaining also a prediction on the non-perturbative physics   
of the type II asymmetric models. On the other hand, we
observed a perturbative super-Higgs mechanism on the type II
asymmetric models, unobservable on the perturbative type IIA duals, 
because it involves as super-Higgs field a modulus 
dual to the type IIA dilaton.

\vskip 1.cm \centerline{\bf Acknowledgements} \noindent The 
authors thank P.M. Petropoulos for valuable discussions. 
A. Gregori thanks the Swiss National Science Foundation, the 
Swiss Office for Education and Science (ofes 95.0856) 
and the EEC, under the contract TMR-ERBFMRX-CT96-0045,
for financial support. J. Rizos thanks 
the CERN Theory Division for hospitality and acknowledges 
financial support from the EEC contract TMR-ERBFMRX-CT96-0090.

\newpage

\noindent

\vskip 0.3cm
\setcounter{section}{0}
\setcounter{equation}{0}
\renewcommand{\theequation}{A.\arabic{equation}}
\section*{\normalsize{\centerline{\bf Appendix A: The type II string
in the free fermionic formulation}}}
In the free fermionic formulation of string theory 
\cite{klt}--\cite{ab} the string 
degrees of freedom are expressed in terms of free world-sheet fermions. 
For the four-dimensional type II string and 
in the light cone-gauge
theory these fermions are \cite{fk}:
\begin{itemize}
\item the space-time degrees of freedom \\[-.7cm]
\begin{flushright}
\begin{minipage}{5.6cm}
left : $\partial_z X_{\mu}(z,{\bar z}),\,\psi^{\rm L}_{\mu}(z)$\\
right :$\partial_{\bar z} X_{\mu}(z,{\bar z}),\,\psi^{\rm R}_{\mu}(z)$\\
\end{minipage}
$\ ,\ \mu=1,2$\hskip3cm\\
\end{flushright}
\item the internal degrees of freedom\\[-.7cm]
\begin{flushright}
\begin{minipage}{5.cm}
left : $\chi^{\rm L}_I(z),\,y^{\rm L}_I(z),\,\omega^{\rm L}_I(z)$\\
right : $\chi^{\rm R}_I(z),\,y^{\rm R}_I(z),\,\omega^{\rm R}_I(z)$\\
\end{minipage}
$\ ,\ I=1,\dots6.$\hskip3cm\\
\end{flushright}
\end{itemize}
The world-sheet supersymmetry, necessary for a consistent theory, 
is realized in the usual way among the space-time coordinates 
and non-linearly among the internal coordinates \cite{abkw}
$$
\delta f^A=\eta^{ABC} f^B f^C \epsilon\ , 
~~~~~~~f\in\{\chi^{\rm L}_I,y^{\rm L}_I,
\omega^{\rm L}_I \,,I=1\dots6\} \, ,
$$
where $\epsilon$ is a Grassmann field and $\eta^{ABC}$ are the properly 
normalised structure constants of a Lie algebra $G=G^L_W$ and similarly
for right movers $G=G^R_W$.

It is known that the 
transportation properties of fermionic fields 
on  surfaces with non-trivial topology, such as the string world-sheet,
are not completely determined by the two-dimensional metric 
and an extra information must be supplied.  This information is known 
as spin structure and describes  the  phases emerging when 
each fermionic field moves around a non-contractible  circle of the 
surface.  Spin structures are in principle arbitrary, but a consistent 
string model should  satisfy a set of physical requirements:
\newcounter{sec}\setcounter{sec}{0}
\begin{list}{{\addtocounter{sec}{1}(\roman{sec})}}{}
\item multiloop modular invariance,
\item factorisation of physical amplitudes,
\item global existence of left and right supercurrents.
\end{list}
After imposing these constraints much freedom is left in choosing spin 
structures and a very big number of consistent string models can be obtained.

If we restrict ourselves to periodic--antiperiodic fermionic fields, 
and demand space-time supersymmetry  to emerge symmetrically from 
left and right movers (which means that left and right space-time fermions will
be treated symmetrically),
the choice of the internal fermion gauge groups 
$(G^L_W,G^R_W)$ is essentially unique: 
$$G^L_W=G^R_W=\prod_{I=1}^6{SO(3)}^I$$
and the left and right supercurrents take the form
\be
T_F(z)=i\partial_z X^\mu\psi^{\rm L}_\mu 
+\sum_{I=1}^6\chi^{\rm L}_I y^{\rm L}_I\omega^{\rm L}_I\ , ~~~~~~ 
{\overline T}_F(z)=i\partial_{\bar z} X^\mu \psi^{\rm R}_{\mu} +
\sum_{I=1}^6 \chi^{\rm R}_I y^{\rm R}_I \omega^{\rm R}_I,
\label{TF}
\ee
Then the solution of the consistency constraints can be expressed in 
a simple set of rules for constructing any type II string model. 
A specific model is defined by 
\newcounter{mo}\setcounter{mo}{0}
\begin{list}{{\addtocounter{mo}{1}\arabic{mo})}}{}
\item A set of boundary condition basis vectors 
$\{b_1=1,\dots,b_{N}\}$, which generate a set of $2^N$ boundary conditions
vectors $\Xi$\,:\,
$\xi\in\Xi\leftrightarrow\xi=\sum_{i=1}^{N}m_i b_i, m_i=0,1$.
Each boundary condition vector $b_i$ has  40 
entries\footnote{Along the paper we employed a 
slightly different notation, that is to 
present  a basis vector as the set of periodic fermions. For example,
$b=\{1,1,1,1,1,1,1,0,\dots,0\}$ is written as
$b=\{\psi^{\rm L}_{\mu},\chi^{\rm L}_1,\chi^{\rm L}_2,\chi^{\rm L}_3,
\chi^{\rm L}_4,\chi^{\rm L}_5,\chi^{\rm L}_6\}$ in this notation.}:
$$
b_i  \equiv \{b_i(\psi^\mu),b_i(\chi^1),\dots,b_i(y^1)\dots,b_i(\omega^1)
\dots ;
b_i({\bar\psi^\mu}),b_i(\chi^1),\dots,b_i({\bar y}^1)\dots,
b_i({\bar\omega}^1)\dots\}
$$
where $b_i(f)=0,1$ 
correspond to the transportation properties  of each  fermion 
$f\to-{\rm e}^{i\pi b_i(f)}$. The basis vectors are subject to some 
restrictions, namely
\newcommand{\mod}{{\rm mod\,\,}}
\begin{enumerate}
\item $b_i\cdot b_i=\mod8\,, \forall\ i =0,\dots,N$,
\item $b_i\cdot b_j=\mod4\,, \forall\ i\ne 0 =1,\dots,N$,
\item $ \prod_f b_i(f)b_j(f)b_k(f)b_{\ell}(f)=0\ \mod2\,, \forall\ 
i{\ne}j{\ne}k{\ne} \ell =0,\dots,N$,
\item {$|\{b_i(\chi_I),b_i(y_I),b_i(\omega_I)\}|^2\ =
b_i(\psi^\mu)\,\mod2$ \ , \\ 
$\left|\{b_i(\bar\chi_I),b_i(\bar y_I),
b_i(\bar\omega_I)\}\right|^2=b_i(\bar\psi^\mu)\,\mod2\,$, \,
 $\forall  \ a=1,\dots,6$  and $\forall\ i =0,\dots,N$}.
\end{enumerate}
\item A set of $\frac{N(N-1)}{2}+1$ 
phases $c{1\atopwithdelims[]1}=\pm1\,,c{b_i\atopwithdelims[]b_j}
=\pm1$, $i>j$ (we will use also the notation 
$C_{(b_i|b_j)} \equiv c{b_i\atopwithdelims[]b_j}$),  
which  determine 
the weight of each spin structure to the string partition function. 
\end{list}
The model's partition function is then given by
$$
Z=
\frac{1}{2^N}
\int_{\cal F}\frac{d^2\tau}{\tau_2^3}\frac{1}{|\eta|^4}
\sum_{\alpha,\beta\in\Xi}c{\alpha\atopwithdelims[]\beta}
Z{\alpha\atopwithdelims[]\beta} \, ,
$$
where
$$
Z{\alpha\atopwithdelims[]\beta}=
\prod_{f={\rm left}}
\left(\frac{\vartheta{\alpha(f)\atopwithdelims[]\beta(f)}}{\eta}\right)
^{\frac{1}{2}}
\prod_{f={\rm right}}
\left(\frac{\bar\vartheta{\alpha(f)\atopwithdelims[]\beta(f)}}{\bar\eta}\right)^
{\frac{1}{2}}
$$
and
$c{\alpha\atopwithdelims[]\beta}$ can be expressed in terms of 
$c{b_i\atopwithdelims[]b_j}$, $i>j$ and $c{b_1\atopwithdelims[]b_1}\equiv
c{1\atopwithdelims[]1}$
using
\ba
c{0\atopwithdelims[]\alpha}&=&\delta_\alpha\\
c{\alpha\atopwithdelims[]{\beta+\gamma}}&=&\delta_\alpha 
c{\alpha\atopwithdelims[]\beta} c{\alpha\atopwithdelims[]\gamma}\\
c{\alpha\atopwithdelims[]\beta}&=& {\rm e}^{\frac{i\pi}{2}\alpha\cdot\beta}
c{\beta\atopwithdelims[]\alpha}\\
c{\alpha\atopwithdelims[]\alpha}&=& {\rm e}^{\frac{i\pi}{4}\alpha^2}
c{\alpha\atopwithdelims[]{b_1}}\\
\ea
with $\delta_\alpha={\rm e}^{i\pi\left(\alpha(\psi^\mu)
+\alpha(\bar\psi^\mu)\right)}$.

Along the paper, we have used indifferently two equivalent notations
for the operation of composition of fermion sets: the sum, as here,
which is more convenient when we specify the fermion sets through
the boundary conditions they assign, 
and the ``symmetric difference'' (see \cite{abk}), indicated as
a product of sets, which contains the union of two fermion sets
minus their intersection, and is more convenient when, as in many sections
of the paper, we specify the periodic fermions contained in the various 
sets.

\noindent

\vskip 0.3cm
\setcounter{section}{0}
\setcounter{equation}{0}
\renewcommand{\theequation}{B.\arabic{equation}}
\section*{\normalsize{\centerline{\bf Appendix B: Massless spectrum and 
vertex operators}}}

\subsection*{\normalsize{\sl B.1.
$Z_2 \times Z_2$ orbifolds in the fermionic construction  }}

We write here the vertex operator representation of the massless states of
the $Z_2 \times Z_2$ orbifold constructed in Section 2.
The massless states of the untwisted sector come from the 
NS-NS, R-NS, NS-R and RR sectors, which are given, in our convention, by 
the sectors $\oslash$
(the ``vacuum''), $S$, $\bar{S}$ and $S\bar{S}$ respectively.
In order to describe the physical states, we
introduce the $SU(2)$,  R-parity currents  ${\rm e}^{\pm {i \over 2}H_i}$,
$\partial H_i$, ${\rm e}^{\pm {i \over 2}\overline{H}_i}$,
$\overline{\partial H}_i$, $i=0,1,2,3$  
\footnote{In this notation, $\partial H \equiv \partial_z  H$,
$\overline{\partial H} \equiv \partial_{\bar{z}}  \overline{H}$,
where $H \equiv X^{\rm L}(z)$,  $\overline{H} \equiv X^{\rm R} (\bar{z})$
are respectively the holomorphic (left-moving) and the
antiholomorphic (right-moving) part into which a world-sheet boson
$X$ is decomposed: $X=H+\overline{H}$.}, where:
\ba
&& \partial  H_0 = \psi^{\rm L}_1 \psi^{\rm L}_2~,~~~
\partial  H_1 = \chi^{\rm L}_1 \chi^{\rm L}_2~,~~~
\partial  H_2 = \chi^{\rm L}_3 \chi^{\rm L}_4~,~~~
\partial  H_3 = \chi^{\rm L}_5 \chi^{\rm L}_6~, \nn \\
&& \nn \\
&& \overline{\partial  H}_0 = \psi^{\rm R}_1 \psi^{\rm R}_2~,~~ \,
\overline{\partial  H}_1 = \chi^{\rm R}_1 \chi^{\rm R}_2~,~~ \,
\overline{\partial  H}_2 = \chi^{\rm R}_3 \chi^{\rm R}_4~,~~ \,
\overline{\partial  H}_3 = \chi^{\rm R}_5 \chi^{\rm R}_6
\label{cur}
\ea
(the eigenvalues of $\partial H_0$  and  $\overline{\partial H}_0$
give the space-time spin).
The fermions originate from the $S$ and $\bar{S}$ sectors.
They are two gravitinos, represented, in the $-1/2$-picture, by
\be
{\rm e}^{\pm {i \over 2}(2H_0+\bar{H}_0+\bar{H}_1+\bar{H}_2+\bar{H}_3)}~,
~~~~~
{\rm e}^{\pm {i \over 2}(2\bar{H}_0+H_0+H_1+H_2+H_3)}~;
\label{gravi}
\ee
two dilatinos:
\be
{\rm e}^{\pm {i \over 2}(2H_0-\bar{H}_0-\bar{H}_1-\bar{H}_2-\bar{H}_3)}~,
~~~~~
{\rm e}^{\pm {i \over 2}(2\bar{H}_0-H_0-H_1-H_2-H_3)}~;
\ee
and twelve spin-1/2 fermions:
\be
{\rm e}^{\pm iH_1}{\rm e}^{\pm {i \over 2} 
(\bar{H}_0+\bar{H}_1-\bar{H}_2-\bar{H}_3)},
~~~~~~
{\rm e}^{\pm i\bar{H}_1}{\rm e}^{\pm {i \over 2} (H_0+H_1-H_2-H_3)},
\ee
\be
{\rm e}^{\pm iH_2}{\rm e}^{\pm {i \over 2} 
(\bar{H}_0-\bar{H}_1+\bar{H}_2-\bar{H}_3)},
~~~~~~
{\rm e}^{\pm i\bar{H}_2}{\rm e}^{\pm {i \over 2} (H_0-H_1+H_2-H_3)},
\ee
\be
{\rm e}^{\pm iH_3}{\rm e}^{\pm {i \over 2} 
(\bar{H}_0-\bar{H}_1-\bar{H}_2+\bar{H}_3)},
~~~~~~
{\rm e}^{\pm i\bar{H}_3}{\rm e}^{\pm {i \over 2} (H_0-H_1-H_2+H_3)},
\ee
The bosons originate from the sectors $\oslash$ and $S\bar{S}$.
The vacuum sector $\oslash$
contains the graviton ${\rm e}^{\pm i (H_0+\bar{H}_0)}$, the pair 
dilaton--pseudoscalar ${\rm e}^{\pm i (H_0-\bar{H}_0)}$, and the six 
complex scalars
${\rm e}^{\pm i (H_i \pm \bar{H}_i)}$.
{}From the vertex operator representation of the graviton and 
the gravitinos we read off
the generators of the $N=2$ supersymmetry:
\be
Q={\rm e}^{-{i \over 2} (H_0 - H_1 - H_2 - H_3)}~~~~{\rm and}~~~~
\bar{Q}={\rm e}^{-{i \over 2} 
(\bar{H}_0 - \bar{H}_1 - \bar{H}_2 - \bar{H}_3)}~,
\label{q1}
\ee
from which we derive also the expressions of the
generators of the $N=2$, $SU(2)$ algebra:
$J^+$, $J^-~~( \equiv J^{+*})$ and  $J^0$, whose charge operator  
$I^0$ is given by:
\be
I^0={1 \over 2}\oint ( \p H_0-\p H_1-\p H_2-\p H_3)- {1 \over 2}\oint
( \overline{\p H_0}-\overline{\p H_1}-\overline{\p H_2}-\overline{\p H_3}).
\label{I0}
\ee
It is then easy to see that the dilaton-pseudoscalar
\be
{\rm e}^{\pm i (H_0-\bar{H}_0)}~,
\label{dil}
\ee
and the three complex scalars
\be
{\rm e}^{\pm i (H_i-\bar{H}_i)}~~~~~~~i=1,2,3,
\label{oscal}
\ee
carry a non-zero $I^0$ charge: they therefore belong to
hypermultiplets. 
Since we are considering a type IIA orbifold, 
the three complex scalars (\ref{oscal})
correspond to the complex structure moduli $U^i$ of the three tori of the 
internal space. The three complex scalars
\be
{\rm e}^{\pm i (H_i+\bar{H}_i)}~~~~~~~i=1,2,3,
\label{ovec}
\ee
have no $I^0$ charge and are superpartners of the vectors: they therefore
correspond to the three K{\"a}hler class moduli $T^i$. 
The $S\bar{S}$ sector contains four complex scalars, which carry $I^0$ charge:
\ba
&& {\rm e}^{\pm {i \over 2}(H_0+H_1+H_2+H_3-\bar{H}_0
-\bar{H}_1-\bar{H}_2-\bar{H}_3)}~,
\nonumber \\
&& {\rm e}^{\pm {i \over 2}(H_0-H_1-H_2+H_3
-\bar{H}_0+\bar{H}_1+\bar{H}_2-\bar{H}_3)}~,
\nonumber \\
&& {\rm e}^{\pm {i \over 2}(H_0-H_1+H_2-H_3
-\bar{H}_0+\bar{H}_1-\bar{H}_2+\bar{H}_3)}~,
\nonumber \\
&& {\rm e}^{\pm {i \over 2}(H_0+H_1-H_2-H_3
-\bar{H}_0-\bar{H}_1+\bar{H}_2+\bar{H}_3)}~,
\label{scal}
\ea
and four vectors:
\ba
&& {\rm e}^{\pm {i \over 2}(H_0+H_1+H_2+H_3
+\bar{H}_0+\bar{H}_1+\bar{H}_2+\bar{H}_3)}~,
\nonumber \\
&& {\rm e}^{\pm {i \over 2}(H_0-H_1-H_2+H_3
+\bar{H}_0-\bar{H}_1-\bar{H}_2+\bar{H}_3)}~,
\nonumber \\
&& {\rm e}^{\pm {i \over 2}(H_0-H_1+H_2-H_3
+\bar{H}_0-\bar{H}_1+\bar{H}_2-\bar{H}_3)}~,
\nonumber \\
&& {\rm e}^{\pm {i \over 2}(H_0+H_1-H_2-H_3
+\bar{H}_0+\bar{H}_1-\bar{H}_2-\bar{H}_3)}~.
\label{vec}
\ea
It is easy to recognise that the first vector belongs to the gravity 
multiplet, being obtained by applying twice the supersymmetry generators
to the graviton. 

\subsection*{\normalsize{\sl B.2.
IIA versus IIB in the fermionic language  }}

The type IIA$\leftrightarrow$B 
exchange is realized by changing the chirality of,
say, the right-moving spinors. In the fermionic construction this is
implemented by the following changes
with respect to the type IIA choice of Table 2.1: 
$C_{(\bar{S}|\bar{S})} \to +1$,
$C_{(\bar{S}|F)} \to -1$, $C_{(\bar{S}|b_{11})} \to -1$, 
$C_{(\bar{S}|b_{22})} \to -1$.
Under this exchange, the right-moving supersymmetry generator becomes:
\be
\bar{Q} \to 
{\rm e}^{-{i \over 2} (\bar{H}_0 + \bar{H}_1 + \bar{H}_2 + \bar{H}_3)}~.
\ee
As a consequence, the $SU(2)$ Cartan charge operator $I^0$ is now:
\be
I^0={1 \over 2} \oint( \p H_0-\p H_1-\p H_2-\p H_3)- {1 \over 2}\oint
( \overline{\p H_0}+\overline{\p H_1}+\overline{\p H_2}+\overline{\p H_3})~.
\ee
In this case the 
states that are associated, in the type IIA construction,
to the complex scalars $T^1$, $T^2$, $T^3$, given in (\ref{ovec}), now
carry $I^0$ charge, while those
associated to the three complex scalars $U^1$, $U^2$, $U^3$,
given in (\ref{oscal}), do not.
The role of $T^i$ and $U^i$ is therefore exchanged, as expected.
A change in the sign of
$C_{(b_{11}|b_{22})}$ exchanges $N_V$ and $N_H$. This then changes 
the sign of the Euler characteristic of the compact space,
$\c=2(N_H-N_V)$.

\subsection*{\normalsize{\sl B.3.
Massless states of type II asymmetric orbifolds }}

The analysis of the massless spectrum of the asymmetric orbifolds
discussed in Section 5 can be performed in a similar 
way, by going to the fermionic point.
We quote here the vertices for 
the bosonic massless states of the untwisted sectors:
$(H^{\rm F},G^{\rm F})$, $(H^{\rm o},G^{\rm o})$ equal to
$(0,0)$ or $(0,1)$.
They are:
\be
\begin{array}{ll}
{\rm e}^{\pm i (H_0+\bar{H}_0)} &   {\rm (graviton)} \\
{\rm e}^{\pm i (H_0-\bar{H}_0)} &   {\rm (dilaton, pseudoscalar)} \\
{\rm e}^{\pm i  (H_i-\bar{H}_j)},~~~i,j=1,2 &  
 {\rm (hyperscalars)} \\
{\rm e}^{\pm i H_0}{\rm e}^{\pm {i \over 2} \bar{H}_3},~~~
{\rm e}^{\pm i  H_3 } {\rm e}^{\pm i \bar{H}_0}~~~ &
{\rm (graviphoton, vectors)} \\
{\rm e}^{\pm i  (H_3-\bar{H}_3)} &  {\rm (vector scalars)} 
\end{array}
\ee
In particular, we notice that, although the dilaton
is represented by the same vertex operator as the dilaton of the
type IIA orbifolds, Eq. (\ref{dil}), in this case
it is uncharged under the $SU(2)$ symmetry of the $N=2$ superalgebra.
The analogous of the operators $Q$ and $\bar{Q}$, Eq. (\ref{q1}), are in fact
in this case:
\be
Q_+,\, Q_-={\rm e}^{-{i \over 2} (H_0 + \epsilon H_1 \pm H_2 \pm H_3)}~,
\label{Qas}
\ee
where $\epsilon$ takes the values $\pm 1$ and depends on the (immaterial)
choice of the chirality of the spinors.
The analogous of the operator $I^0$ is therefore
\be
I^0_{\rm As}= \oint \left( \partial H_2 + \partial H_3 \right)~.
\ee

\noindent

\vskip 0.3cm
\setcounter{section}{0}
\setcounter{equation}{0}
\renewcommand{\theequation}{C.\arabic{equation}}
\section*{\normalsize{\centerline{\bf Appendix C: GSO projections on the
twisted sectors}}}

We quote here the contribution to the quantity
$4(N_V-N_H)$, as a function of the modular
coefficients, of each one of the 48 twisted supersectors.
By supersector we mean a twisted sector and the sectors
derived from it by addition of the sets $S$, $\bar{S}$, $S \bar{S}$,
which provide the supersymmetric partners.
For simplicity, we omit to indicate the latter,
so that $b_1$ is a short-hand notation for the sets $b_1$, $Sb_1$, 
$\bar{S}b_1$, $S\bar{S}b_1$.  
The quantity $4(N_V+N_H)$ is given by the product of the two
square brackets.

\underline{$b_{1}$}:
\be
[1+C_{(e_{1}|b_{1})}][1+C_{(e_{2}|b_{1})}]\cdot \alpha,  
\label{sb1}
\ee
\underline{$b_{1}+e_{3}$}:
\be
[1+C_{(e_{1}|b_{1})}C_{(e_{1}|e_{3})}][1+C_{(e_{2}|b_{1})}C_{(e_{2}|e_{3})}]
\cdot \alpha C_{(e_{3}|e_{4})}C_{(e_{3}|b_{2})}, 
\ee
\underline{$b_{1}+e_{4}$}:
\be
[1+C_{(e_{1}|b_{1})}C_{(e_{4}|e_{1})}][1+C_{(e_{2}|b_{1})}C_{(e_{2}|e_{4})}]
\cdot \alpha C_{(e_{3}|e_{4})}C_{(e_{4}|b_{2})}, 
\ee
\underline{$b_{1}+e_{5}$}:
\be
[1+C_{(e_{1}|b_{1})}C_{(e_{1}|e_{5})}][1+C_{(b_{1}|e_{2})}C_{(b_{2}|e_{5})}]
\cdot \alpha \gamma C_{(e_{3}|e_{5})}C_{(e_{4}|e_{5})},
\ee
\underline{$b_{1}+e_{6}$}:
\be
[1+C_{(b_{1}|e_{1})}C_{(e_{1}|e_{6})}][1+C_{(b_{1}|e_{2})}C_{(e_{2}|e_{6})}]
\cdot \gamma \beta C_{(e_{5}|e_{6})}C_{(b_{2}|e_{3})}C_{(b_{2}|e_{4})},
\ee
\underline{$b_{1}+e_{3}+e_{4}$}:
\be
[1+C_{(b_{1}|e_{1})}C_{(e_{1}|e_{3})}C_{(e_{1}|e_{4})}]
[1+C_{(b_{1}|e_{2})}C_{(e_{2}|e_{3})}C_{(e_{2}|e_{4})}]
\cdot \alpha C_{(b_{2}|e_{3})}C_{(b_{2}|e_{4})}, 
\ee
\underline{$b_{1}+e_{3}+e_{5}$}:
\begin{eqnarray}
\lefteqn{
[1+C_{(b_{1}|e_{1})}C_{(e_{1}|e_{3})}C_{(e_{1}|e_{5})}] \times } \\ 
&& \times [1+C_{(b_{1}|e_{2})}C_{(e_{2}|e_{3})}C_{(e_{2}|e_{5})}]
\cdot \alpha \gamma C_{(e_{3}|e_{4})}C_{(e_{4}|e_{5})}C_{(b_{2}|e_{3})},
\nonumber
\end{eqnarray}
\underline{$b_{1}+e_{3}+e_{6}$}:
\begin{eqnarray}
\lefteqn{
[1+C_{(b_{1}|e_{1})}C_{(e_{1}|e_{2})}C_{(e_{1}|e_{4})}C_{(e_{1}|e_{5})}]
\times } \\ 
&& \times 
[1+C_{(b_{1}|e_{2})}C_{(e_{1}|e_{2})}C_{(e_{2}|e_{4})}C_{(e_{2}|e_{5})}]
\cdot \beta \gamma C_{(b_{2}|e_{4})}C_{(e_{1}|e_{5})}C_{(e_{2}|e_{5})}
C_{(e_{4}|e_{5})},
\nonumber
\end{eqnarray}
\underline{$b_{1}+e_{4}+e_{5}$}:
\begin{eqnarray}
\lefteqn{
[1+C_{(b_{1}|e_{1})}C_{(e_{1}|e_{4})}C_{(e_{1}|e_{5})}] \times } \\
&& \times [1+C_{(b_{1}|e_{2})}C_{(e_{2}|e_{4})}C_{(e_{2}|e_{5})}]
\cdot \alpha \gamma C_{(b_{2}|e_{4})}C_{(e_{3}|e_{4})}C_{(e_{3}|e_{5})},
\nonumber
\end{eqnarray}
\underline{$b_{1}+e_{4}+e_{6}$}:
\begin{eqnarray}
\lefteqn{
[1+C_{(b_{1}|e_{1})}C_{(e_{1}|e_{2})}C_{(e_{1}|e_{3})}C_{(e_{1}|e_{5})}]
\times } \\ 
&& \times
[1+C_{(b_{1}|e_{2})}C_{(e_{1}|e_{2})}C_{(e_{2}|e_{3})}C_{(e_{2}|e_{5})}]
\cdot \beta \gamma C_{(b_{2}|e_{3})}C_{(e_{1}|e_{5})}
C_{(e_{2}|e_{5})}C_{(e_{3}|e_{5})},
\nonumber
\end{eqnarray}
\underline{$b_{1}+e_{5}+e_{6}$}:
\begin{eqnarray}
\lefteqn{
[1+C_{(b_{1}|e_{1})}C_{(e_{1}|e_{2})}C_{(e_{1}|e_{3})}C_{(e_{1}|e_{4})}]
\times } \\
&& \times 
[1+C_{(b_{1}|e_{2})}C_{(e_{1}|e_{2})}C_{(e_{2}|e_{3})}C_{(e_{2}|e_{4})}]
\cdot \beta  C_{(b_{2}|e_{3})}C_{(b_{2}|e_{4})},
\nonumber
\end{eqnarray}
\underline{$b_{1}+e_{4}+e_{5}+e_{6}$}:
\be
[1+C_{(b_{1}|e_{1})}C_{(e_{1}|e_{2})}C_{(e_{1}|e_{3})}]
[1+C_{(b_{1}|e_{2})}C_{(e_{1}|e_{2})}C_{(e_{2}|e_{3})}]
\cdot \beta C_{(b_{2}|e_{3})},
\ee
\underline{$b_{1}+e_{3}+e_{4}+e_{6}$}:
\be
[1+C_{(b_{1}|e_{1})}C_{(e_{1}|e_{2})}C_{(e_{1}|e_{5})}]
[1+C_{(b_{1}|e_{2})}C_{(e_{1}|e_{2})}C_{(e_{2}|e_{5})}]
\cdot \beta \gamma C_{(e_{1}|e_{5})}
C_{(e_{2}|e_{5})},
\ee
\underline{$b_{1}+e_{3}+e_{5}+e_{6}$}:
\be
[1+C_{(b_{1}|e_{1})}C_{(e_{1}|e_{2})}C_{(e_{1}|e_{4})}]
[1+C_{(b_{1}|e_{2})}C_{(e_{1}|e_{2})}C_{(e_{2}|e_{4})}]
\cdot \beta  C_{(b_{2}|e_{4})},
\ee
\underline{$b_{1}+e_{3}+e_{4}+e_{5}$}:
\begin{eqnarray}
\lefteqn{
[1+C_{(b_{1}|e_{1})}C_{(e_{1}|e_{3})}C_{(e_{1}|e_{4})}C_{(e_{1}|e_{5})}]
\times} \\ 
&& \times
[1+C_{(b_{1}|e_{2})}C_{(e_{2}|e_{3})}C_{(e_{2}|e_{4})}C_{(e_{2}|e_{5})}]
\cdot \alpha \gamma C_{(b_{2}|e_{3})}C_{(b_{2}|e_{4})},
\nonumber
\end{eqnarray}
\underline{$b_{1}+e_{3}+e_{4}+e_{5}+e_{6}$}:
\be
[1+C_{(b_{1}|e_{1})}C_{(e_{1}|e_{2})}][1+C_{(b_{1}|e_{2})}C_{(e_{1}|e_{2})}]
\cdot \beta, 
\ee
\underline{$b_{2}$}:
\be
[1+C_{(b_{2}|e_{3})}][1+C_{(b_{2}|e_{4})}] \cdot \beta, 
\ee
\underline{$b_{2}+e_{1}$}:
\be
[1+C_{(b_{2}|e_{3})}C_{(e_{1}|e_{3})}]
[1+C_{(b_{2}|e_{4})}C_{(e_{1}|e_{4})}] 
\cdot \beta C_{(b_{1}|e_{1})}C_{(e_{1}|e_{2})}, 
\ee
\underline{$b_{2}+e_{2}$}:
\be
[1+C_{(b_{2}|e_{3})}C_{(e_{2}|e_{3})}]
[1+C_{(b_{2}|e_{4})}C_{(e_{2}|e_{4})}] 
\cdot \beta C_{(b_{1}|e_{2})}C_{(e_{1}|e_{2})}, 
\ee
\underline{$b_{2}+e_{5}$}:
\be
[1+C_{(b_{2}|e_{3})}C_{(e_{3}|e_{5})}]
[1+C_{(b_{2}|e_{4})}C_{(e_{4}|e_{5})}] 
\cdot \beta \gamma C_{(e_{1}|e_{5})}C_{(e_{2}|e_{5})}, 
\ee
\underline{$b_{2}+e_{6}$}:
\begin{eqnarray}
\lefteqn{
[1+C_{(b_{2}|e_{3})}C_{(e_{1}|e_{3})}C_{(e_{2}|e_{3})}C_{(e_{4}|e_{3})}
C_{(e_{5}|e_{3})}]
[1+C_{(b_{2}|e_{4})}C_{(e_{1}|e_{4})}C_{(e_{2}|e_{4})}
\cdot } \\ 
&& \cdot C_{(e_{3}|e_{4})}C_{(e_{5}|e_{4})}] 
\cdot \alpha \gamma C_{(b_{1}|e_{1})}C_{(b_{1}|e_{2})}C_{(e_{1}|e_{5})}
C_{(e_{2}|e_{5})}C_{(e_{3}|e_{5})}C_{(e_{4}|e_{5})}, 
\nonumber
\end{eqnarray}
\underline{$b_{2}+e_{1}+e_{2}$}:
\be
[1+C_{(b_{2}|e_{3})}C_{(e_{1}|e_{3})}C_{(e_{2}|e_{3})}]
[1+C_{(b_{2}|e_{4})}C_{(e_{1}|e_{4})}C_{(e_{2}|e_{4})}] 
\cdot \beta C_{(b_{1}|e_{1})}C_{(b_{1}|e_{2})}, 
\ee
\underline{$b_{2}+e_{1}+e_{5}$}:
\begin{eqnarray}
\lefteqn{
[1+C_{(b_{2}|e_{3})}C_{(e_{1}|e_{3})}C_{(e_{5}|e_{3})}] \times } \\
&& \times [1+C_{(b_{2}|e_{4})}C_{(e_{1}|e_{4})}C_{(e_{5}|e_{4})}] 
\cdot \beta \gamma C_{(b_{1}|e_{1})}C_{(e_{1}|e_{2})}
C_{(e_{5}|e_{2})},
\nonumber
\end{eqnarray}
\underline{$b_{2}+e_{2}+e_{5}$}:
\begin{eqnarray}
\lefteqn{
[1+C_{(b_{2}|e_{3})}C_{(e_{2}|e_{3})}C_{(e_{5}|e_{3})}] \times } \\ 
&& \times [1+C_{(b_{2}|e_{4})}C_{(e_{1}|e_{4})}C_{(e_{5}|e_{4})}] 
\cdot  \beta \gamma C_{(b_{1}|e_{2})}C_{(e_{1}|e_{2})}C_{(e_{1}|e_{5})}, 
\nonumber
\end{eqnarray}
\underline{$b_{2}+e_{2}+e_{6}$}:
\begin{eqnarray}
\lefteqn{
[1+C_{(b_{2}|e_{3})}C_{(e_{1}|e_{3})}C_{(e_{4}|e_{3})}C_{(e_{5}|e_{3})}]
\times } \\ 
&& \times 
[1+C_{(b_{2}|e_{4})}C_{(e_{1}|e_{4})}C_{(e_{3}|e_{4})}C_{(e_{5}|e_{4})}] 
\cdot \alpha \gamma C_{(b_{1}|e_{1})}C_{(e_{1}|e_{5})}C_{(e_{3}|e_{5})}
C_{(e_{4}|e_{5})}, \nonumber
\end{eqnarray}
\underline{$b_{2}+e_{1}+e_{6}$}:
\begin{eqnarray}
\lefteqn{
[1+C_{(b_{2}|e_{3})}C_{(e_{2}|e_{3})}C_{(e_{4}|e_{3})}C_{(e_{5}|e_{3})}]
\times } \\
&& \times
[1+C_{(b_{2}|e_{4})}C_{(e_{2}|e_{4})}C_{(e_{3}|e_{4})}C_{(e_{5}|e_{4})}] 
\cdot \alpha \gamma C_{(b_{1}|e_{2})}C_{(e_{2}|e_{5})}C_{(e_{3}|e_{5})}
C_{(e_{4}|e_{5})}, \nonumber
\end{eqnarray}
\underline{$b_{2}+e_{5}+e_{6}$}:
\begin{eqnarray}
\lefteqn{
[1+C_{(b_{2}|e_{3})}C_{(e_{1}|e_{3})}C_{(e_{2}|e_{3})}C_{(e_{4}|e_{3})}]
\times } \\ 
&& \times
[1+C_{(b_{2}|e_{4})}C_{(e_{1}|e_{4})}C_{(e_{2}|e_{4})}C_{(e_{3}|e_{4})}] 
\cdot \alpha C_{(b_{1}|e_{1})}C_{(b_{1}|e_{2})}, \nonumber
\end{eqnarray}
\underline{$b_{2}+e_{1}+e_{2}+e_{5}$}:
\begin{eqnarray}
\lefteqn{
[1+C_{(b_{2}|e_{3})}C_{(e_{1}|e_{3})}C_{(e_{2}|e_{3})}C_{(e_{5}|e_{3})}]
\times } \\ 
&& \times
[1+C_{(b_{2}|e_{4})}C_{(e_{1}|e_{4})}C_{(e_{2}|e_{4})}C_{(e_{5}|e_{4})}] 
\cdot \beta \gamma C_{(b_{1}|e_{1})}C_{(b_{1}|e_{2})},
\nonumber
\end{eqnarray}
\underline{$b_{2}+e_{1}+e_{2}+e_{6}$}:
\be
[1+C_{(b_{2}|e_{3})}C_{(e_{4}|e_{3})}C_{(e_{5}|e_{3})}]
[1+C_{(b_{2}|e_{4})}C_{(e_{3}|e_{4})}C_{(e_{5}|e_{4})}] 
\cdot \alpha \gamma C_{(e_{3}|e_{5})}C_{(e_{4}|e_{5})},
\ee
\underline{$b_{2}+e_{1}+e_{5}+e_{6}$}:
\be
[1+C_{(b_{2}|e_{3})}C_{(e_{2}|e_{3})}C_{(e_{4}|e_{3})}]
[1+C_{(b_{2}|e_{4})}C_{(e_{2}|e_{4})}C_{(e_{3}|e_{4})}] 
\cdot \alpha C_{(b_{1}|e_{2})},
\ee
\underline{$b_{2}+e_{2}+e_{5}+e_{6}$}:
\be
[1+C_{(b_{2}|e_{3})}C_{(e_{1}|e_{3})}C_{(e_{4}|e_{3})}]
[1+C_{(b_{2}|e_{4})}C_{(e_{1}|e_{4})}C_{(e_{3}|e_{4})}] 
\cdot \alpha C_{(b_{1}|e_{1})},
\ee
\underline{$b_{2}+e_{1}+e_{2}+e_{5}+e_{6}$}:
\be
[1+C_{(b_{2}|e_{3})}C_{(e_{4}|e_{3})}]
[1+C_{(b_{2}|e_{4})}C_{(e_{3}|e_{4})}] 
\cdot \alpha,
\ee
\underline{$b_{3}$}:
\be
[1+\gamma]
[1+\alpha \beta C_{(b_{1}|e_{1})}C_{(b_{1}|e_{2})}C_{(b_{2}|e_{3})}
C_{(b_{2}|e_{4})}] 
\cdot \beta C_{(b_{1}|e_{1})}C_{(b_{1}|e_{2})}, 
\ee
\underline{$b_{3}+e_{1}$}:
\begin{eqnarray}
\lefteqn{
[1+\gamma C_{(e_{1}|e_{5})}]
[1+\alpha \beta C_{(b_{1}|e_{1})}C_{(b_{1}|e_{2})}C_{(b_{2}|e_{3})}
\cdot } \\
&& \cdot
C_{(b_{2}|e_{4})}C_{(e_{1}|e_{2})}C_{(e_{1}|e_{3})}C_{(e_{1}|e_{4})}] 
\cdot \beta C_{(b_{1}|e_{2})}C_{(e_{1}|e_{2})}, 
\nonumber
\end{eqnarray}
\underline{$b_{3}+e_{2}$}:
\begin{eqnarray}
\lefteqn{
[1+\gamma C_{(e_{2}|e_{5})}]
[1+\alpha \beta \gamma C_{(b_{1}|e_{1})}C_{(b_{1}|e_{2})}C_{(b_{2}|e_{3})}
C_{(b_{2}|e_{4})}C_{(e_{1}|e_{2})} \cdot } \\ 
&& \cdot C_{(e_{2}|e_{3})}C_{(e_{2}|e_{4})}
C_{(e_{2}|e_{5})}] 
\cdot \beta C_{(b_{1}|e_{1})}C_{(e_{1}|e_{2})}, \nonumber
\end{eqnarray}
\underline{$b_{3}+e_{3}$}:
\begin{eqnarray}
\lefteqn{
[1+\gamma C_{(e_{3}|e_{5})}]
[1+\alpha \beta \gamma C_{(b_{1}|e_{1})}C_{(b_{1}|e_{2})}C_{(b_{2}|e_{3})}
C_{(b_{2}|e_{4})}C_{(e_{3}|e_{1})} \cdot } \\ 
&& \cdot C_{(e_{3}|e_{2})}C_{(e_{3}|e_{4})}
C_{(e_{3}|e_{5})}] 
\cdot \alpha C_{(b_{2}|e_{4})}C_{(e_{3}|e_{4})}, \nonumber
\end{eqnarray}
\underline{$b_{3}+e_{4}$}:
\begin{eqnarray}
\lefteqn{
[1+\gamma C_{(e_{4}|e_{5})}]
[1+\alpha \beta \gamma C_{(b_{1}|e_{1})}C_{(b_{1}|e_{2})}C_{(b_{2}|e_{3})}
C_{(b_{2}|e_{4})}C_{(e_{4}|e_{1})} \cdot } \\ 
&& \cdot C_{(e_{4}|e_{2})}C_{(e_{4}|e_{3})}
C_{(e_{4}|e_{5})}] 
\cdot \alpha C_{(b_{2}|e_{3})}C_{(e_{4}|e_{3})}, \nonumber
\end{eqnarray}
\underline{$b_{3}+e_{1}+e_{2}$}:
\begin{eqnarray}
\lefteqn{
[1+\gamma C_{(e_{1}|e_{5})}C_{(e_{2}|e_{5})}]
[1+\alpha \beta \gamma C_{(b_{1}|e_{1})}C_{(b_{1}|e_{2})}C_{(b_{2}|e_{3})}
\cdot } \\
&& \cdot
C_{(b_{2}|e_{4})}C_{(e_{1}|e_{3})}C_{(e_{1}|e_{4})} C_{(e_{1}|e_{5})}
C_{(e_{2}|e_{3})}C_{(e_{2}|e_{4})}C_{(e_{2}|e_{5})}] 
\cdot \beta, \nonumber
\end{eqnarray}
\underline{$b_{3}+e_{1}+e_{3}$}:
\begin{eqnarray}
\lefteqn{
[1+\gamma C_{(e_{1}|e_{5})}C_{(e_{3}|e_{5})}]
[1+\alpha \beta \gamma C_{(b_{1}|e_{1})}C_{(b_{1}|e_{2})}C_{(b_{2}|e_{3})}
C_{(b_{2}|e_{4})}C_{(e_{1}|e_{2})}C_{(e_{1}|e_{4})} \cdot } \\
&& \cdot C_{(e_{1}|e_{5})}
C_{(e_{2}|e_{3})}C_{(e_{3}|e_{4})}C_{(e_{3}|e_{5})}] 
\cdot \beta C_{(b_{1}|e_{2})}C_{(b_{2}|e_{3})}C_{(e_{1}|e_{2})}
C_{(e_{2}|e_{3})}, \nonumber
\end{eqnarray}
\underline{$b_{3}+e_{1}+e_{4}$}:
\begin{eqnarray}
\lefteqn{
[1+\gamma C_{(e_{1}|e_{5})}C_{(e_{4}|e_{5})}]
[1+\alpha \beta \gamma C_{(b_{1}|e_{1})}C_{(b_{1}|e_{2})}C_{(b_{2}|e_{3})}
C_{(b_{2}|e_{4})}C_{(e_{1}|e_{2})}C_{(e_{1}|e_{3})} \cdot } \\ 
&& \cdot C_{(e_{1}|e_{5})}
C_{(e_{2}|e_{4})}C_{(e_{3}|e_{4})}C_{(e_{4}|e_{5})}] 
\cdot \beta C_{(b_{1}|e_{2})}C_{(b_{2}|e_{4})}C_{(e_{1}|e_{2})}
C_{(e_{2}|e_{4})}, \nonumber
\end{eqnarray}
\underline{$b_{3}+e_{2}+e_{3}$}:
\begin{eqnarray}
\lefteqn{
[1+\gamma C_{(e_{2}|e_{5})}C_{(e_{3}|e_{5})}]
[1+\alpha \beta \gamma C_{(b_{1}|e_{1})}C_{(b_{1}|e_{2})}C_{(b_{2}|e_{3})}
C_{(b_{2}|e_{4})}C_{(e_{1}|e_{2})}C_{(e_{2}|e_{4})} \cdot } \\ 
&& \cdot C_{(e_{2}|e_{5})}
C_{(e_{1}|e_{3})}C_{(e_{3}|e_{4})}C_{(e_{3}|e_{5})}] 
\cdot \beta C_{(b_{1}|e_{2})}C_{(b_{2}|e_{3})}C_{(e_{1}|e_{2})}
C_{(e_{1}|e_{3})}, \nonumber
\end{eqnarray}
\underline{$b_{3}+e_{2}+e_{4}$}:
\begin{eqnarray}
\lefteqn{
[1+\gamma C_{(e_{2}|e_{5})}C_{(e_{4}|e_{5})}]
[1+\alpha \beta \gamma C_{(b_{1}|e_{1})}C_{(b_{1}|e_{2})}C_{(b_{2}|e_{3})}
C_{(b_{2}|e_{4})}C_{(e_{1}|e_{2})}C_{(e_{2}|e_{3})} \cdot } \\ 
&& \cdot C_{(e_{2}|e_{5})}
C_{(e_{1}|e_{4})}C_{(e_{3}|e_{4})}C_{(e_{4}|e_{5})}] 
\cdot \beta C_{(b_{1}|e_{2})}C_{(b_{2}|e_{4})}C_{(e_{1}|e_{2})}
C_{(e_{1}|e_{4})}, \nonumber
\end{eqnarray}
\underline{$b_{3}+e_{3}+e_{4}$}:
\begin{eqnarray}
\lefteqn{
[1+\gamma C_{(e_{3}|e_{5})}C_{(e_{4}|e_{5})}]
[1+\alpha \beta \gamma C_{(b_{1}|e_{1})}C_{(b_{1}|e_{2})}C_{(b_{2}|e_{3})}
\cdot } \\
&& \cdot C_{(b_{2}|e_{4})}C_{(e_{3}|e_{1})}C_{(e_{3}|e_{2})} 
C_{(e_{3}|e_{5})}C_{(e_{4}|e_{1})}C_{(e_{4}|e_{2})}C_{(e_{4}|e_{5})}] 
\cdot \alpha , \nonumber
\end{eqnarray}
\underline{$b_{3}+e_{1}+e_{2}+e_{3}$}:
\begin{eqnarray}
\lefteqn{
[1+\gamma C_{(e_{1}|e_{5})}C_{(e_{2}|e_{5})}C_{(e_{3}|e_{5})}]
[1+\alpha \beta \gamma C_{(b_{1}|e_{1})}C_{(b_{1}|e_{2})}C_{(b_{2}|e_{3})}
C_{(b_{2}|e_{4})} \cdot } \\ 
&& \cdot C_{(e_{1}|e_{4})}C_{(e_{1}|e_{5})}C_{(e_{2}|e_{4})}
C_{(e_{2}|e_{5})}C_{(e_{3}|e_{4})}C_{(e_{3}|e_{5})}] 
\cdot \beta C_{(b_{2}|e_{3})}, \nonumber
\end{eqnarray}
\underline{$b_{3}+e_{1}+e_{3}+e_{4}$}:
\begin{eqnarray}
\lefteqn{
[1+\gamma C_{(e_{1}|e_{5})}C_{(e_{3}|e_{5})}C_{(e_{4}|e_{5})}]
[1+\alpha \beta \gamma C_{(b_{1}|e_{1})}C_{(b_{1}|e_{2})}C_{(b_{2}|e_{3})}
C_{(b_{2}|e_{4})} \cdot } \\ 
&& \cdot C_{(e_{1}|e_{2})}C_{(e_{1}|e_{5})}C_{(e_{3}|e_{2})}
C_{(e_{3}|e_{5})}C_{(e_{4}|e_{2})}C_{(e_{4}|e_{5})}] 
\cdot \alpha C_{(b_{1}|e_{1})} , \nonumber
\end{eqnarray}
\underline{$b_{3}+e_{1}+e_{2}+e_{4}$}:
\begin{eqnarray}
\lefteqn{
[1+\gamma C_{(e_{1}|e_{5})}C_{(e_{2}|e_{5})}C_{(e_{4}|e_{5})}]
[1+\alpha \beta \gamma C_{(b_{1}|e_{1})}C_{(b_{1}|e_{2})}C_{(b_{2}|e_{3})}
C_{(b_{2}|e_{4})} \cdot } \\ 
&& \cdot C_{(e_{1}|e_{3})}C_{(e_{1}|e_{5})}C_{(e_{2}|e_{3})}
C_{(e_{2}|e_{5})}C_{(e_{4}|e_{3})}C_{(e_{4}|e_{5})}] 
\cdot \beta C_{(b_{2}|e_{4})} , \nonumber
\end{eqnarray}
\underline{$b_{3}+e_{2}+e_{3}+e_{4}$}:
\begin{eqnarray}
\lefteqn{
[1+\gamma C_{(e_{2}|e_{5})}C_{(e_{3}|e_{5})}C_{(e_{4}|e_{5})}]
[1+\alpha \beta \gamma C_{(b_{1}|e_{1})}C_{(b_{1}|e_{2})}C_{(b_{2}|e_{3})}
C_{(b_{2}|e_{4})} \cdot } \\ 
&& \cdot C_{(e_{1}|e_{2})}C_{(e_{2}|e_{5})}C_{(e_{3}|e_{1})}
C_{(e_{3}|e_{5})}C_{(e_{4}|e_{1})}C_{(e_{4}|e_{5})}] 
\cdot \alpha C_{(b_{1}|e_{2})} , \nonumber
\end{eqnarray}
\underline{$b_{3}+e_{1}+e_{2}+e_{3}+e_{4}$}:
\begin{eqnarray}
\lefteqn{
[1+\gamma C_{(e_{1}|e_{5})}C_{(e_{2}|e_{5})}C_{(e_{3}|e_{5})}
C_{(e_{4}|e_{5})}]
[1+\alpha \beta \gamma C_{(b_{1}|e_{1})}C_{(b_{1}|e_{2})}C_{(b_{2}|e_{3})}
\cdot } \label{slast} \\ 
&& \cdot C_{(b_{2}|e_{4})}C_{(e_{1}|e_{5})}C_{(e_{2}|e_{5})}C_{(e_{3}|e_{5})}
C_{(e_{4}|e_{5})}] 
\cdot \alpha C_{b_{1}|e_{1})}C_{(b_{1}|e_{2})} , \nonumber
\end{eqnarray}

In the previous expressions, all the modular 
coefficients are symmetric under the
exchange of the arguments: $C_{(\alpha|\beta)}=
C_{(\beta|\alpha)}$. We defined also:
\begin{eqnarray}
\alpha &=& C_{(b_{1}|F)}C_{(b_{1}|e_{3})}C_{(b_{1}|e_{4})}, \nonumber \\
\beta &=& C_{(b_{2}|F)}C_{(b_{2}|e_{1})}C_{(b_{2}|e_{2})}, \\
\gamma &=& C_{(b_{1}|e_{5})}C_{(b_{2}|e_{5})}. \nonumber 
\end{eqnarray}

By running the above formulae with a computer program, we find the following
pairs of $(N_V,N_H)$: $(0,48)$, $(48,0)$, $(28,4)$, $(4,28)$,  
$(16,16)$, $(0,24)$, $(24,0)$, $(6,18)$, $(18,6)$,
$(12,12)$, $(4,16)$, $(16,4)$, $(2,14)$, $(14,2)$, $(8,8)$,       
$(0,12)$, $(12,0)$,  $(6,6)$,  $(3,9)$, $(9,3)$, $(4,4)$,  $(2,2)$,
$(0,0)$.

The formulae above simplify if, instead of factorizing the boundary 
conditions of the
six circles with the sets $e_i$, $i=1,...,5$, we consider only the
factorization of the three tori $(1,2)$, $(3,4)$ and $(5,6)$ with the sets
$T_i$, $i=1,2,3$, given by:
\ba
T_1 & = & \left\{ y_1^L,y_2^L,\o_1^L,\o_2^L~|~y_1^R,y_2^R,\o_1^R,\o_2^R 
\right\}~, \nonumber \\
T_2 & = & \left\{ y_3^L,y_4^L,\o_3^L,\o_4^L~|~y_3^R,y_4^R,\o_3^R,\o_4^R 
\right\}~, \\
T_3 & = & \left\{ y_5^L,y_6^L,\o_5^L,\o_6^L~|~y_5^R,y_6^R,\o_5^R,\o_6^R 
\right\}~. \nonumber 
\ea
In this case, we can write a compact expression for the formulae, 
which give the sum and the difference of the total number of vector and 
hypermultiplets provided by the twisted sectors:
\ba
\frac{N_{V}+N_{H}}{4} & = & 6+(1+C_{(T_{1}|T_{2})}) \times  \\
&& \;\;\; \times (C_{(b_{1}|T_{1})}+C_{(b_{2}|T_{2})}+ \alpha \beta
C_{(b_{1}|T_{1})}C_{(b_{2}|T_{2})}), \nonumber \\
\frac{N_{V}-N_{H}}{4} & = & \frac{3}{2} C_{(b_{1}|b_{2})} (\alpha + \beta)
\times \\
&& \;\;\; \times (1+C_{(b_{1}|T_{1})} + C_{(b_{2}|T_{2})} + 
C_{(b_{1}|T_{1})}C_{(b_{2}|T_{2})} C_{(T_{1}|T_{2})}), \nonumber
\ea
where $\alpha$ and $\beta$ are given by:
\ba
\alpha & = & C_{(b_{1}|F)}C_{(b_{1}|T_{2})}, \nonumber \\
\beta & = & C_{(b_{2}|F)}C_{(b_{2}|T_{1})}.
\ea 
In this case, we can only find a subset of models, namely those with
$(N_V,N_H)=(48,0)$, $(0,48)$, $(24,0)$, $(0,24)$, $(16,16)$, $(12,12)$, 
$(8,8)$ and $(0,0)$.

\subsection*{\normalsize{\sl C.1. A choice of modular coefficients
for each model  }}

We give here a choice of modular coefficients per each one of the constructions
of Section 5:

1) (48,0). All coefficients $=+1$.
\
\\

2) (28,4). $(D,D,O)$: $C_{(e_1|e_3)}$, $C_{(e_1|e_4)}$, $C_{(e_2|e_3)}$,
$C_{(e_2|e_4)}=-1$. 
\
\\

3) (16,16). $(O,O,F)$: $C_{(b_1|F)}=-1$.
\
\\

4) (24,0). $(D,D,D)$: $C_{(e_2|e_5)}$, $C_{(e_4|e_5)}=-1$.
\
\\

5) (18,6).  $(DD,DD,O)$: $C_{(e_1|e_2)}$, $C_{(e_1|e_4)}$, $C_{(e_2|e_3)}$,
$C_{(e_3|e_4)}=-1$. 
\
\\

6) (12,12). $(R,O,FD)$: $C_{(e_1|e_2)}$, $\gamma=-1$.

$(D,D,D)$: $C_{(e_1|e_2)}$, $C_{(e_3|e_4)}=-1$.
\
\\

7) (16,4). $(DD,D,D)$: $C_{(e_1|e_2)}$, $C_{(e_2|e_4)}=-1$.
\
\\

8) (14,2). $(DD,DD,D)$: $C_{(e_1|e_4)}$, $C_{(e_2|e_3)}=-1$.
\
\\

9) (8,8). $(DD,DD,D)$: $C_{(e_1|e_2)}$, $C_{(e_2|e_4)}$, $C_{(e_3|e_4)}=-1$.

$(D,D,F)$: $C_{(e_1|e_3)}$, $C_{(e_1|e_4)}$, $C_{(e_2|e_3)}$, 
$C_{(e_3|e_4)}$, $\gamma=-1$.
$(D,D,FD)$: $C_{(e_1|e_3)}$, $\gamma=-1$.

$(O,F,F)$: $C_{(b_2|e_3)}$.
$(O,FD,FD)$: $C_{(b_2|e_3)}$, $C_{(e_4|e_5)}=-1$.
\
\\

10) (12,0). $(DD,DD,DD)$: $C_{(e_1|e_5)}$, $C_{(e_2|e_4)}$, $C_{(e_3|e_5)}=-1$.
\
\\

11) (6,6). $(DD,DD,DD)$: $C_{(e_1|e_2)}$, $C_{(e_2|e_5)}$, $C_{(e_3|e_4)}$, 
$C_{(e_4|e_5)}=-1$.

$(DD,D,FD)$: $C_{(e_1|e_2)}$, $C_{(e_2|e_4)}$, $\gamma=-1$.
\
\\

12) (9,3). $(RR,RR,RR)$: $C_{(e_1|e_2)}$, $C_{(e_2|e_4)}$, 
$C_{(e_3|e_4)}$, $C_{(e_3|e_5)}$, $\alpha$, $\beta=-1$.
\
\\

13) (4,4). $(D,FD,F)$: $C_{(b_2|e_3)}$, $C_{(e_1|e_3)}$, $C_{(e_1|e_4)}$, 
$C_{(e_2|e_3)}$, $C_{(e_2|e_4)}=-1$. 
$(D,FD,FD)$: $C_{(b_2|e_3)}$, 
$C_{(e_2|e_5)}$, $C_{(e_4|e_5)}$, $\alpha=-1$.

$(DD,DD,F)$: $C_{(e_1|e_2)}$, $C_{(e_1|e_4)}$, 
$C_{(e_2|e_3)}$, $C_{(e_3|e_4)}$, $\gamma=-1$.
$(DD,DD,FD)$: $C_{(e_1|e_3)}$, $C_{(e_2|e_4)}$, $\gamma=-1$. 
\
\\

14) (2,2). $(DD,FD,FD)$: $C_{(b_2|e_4)}$, $C_{(e_1|e_2)}$, $C_{(e_2|e_3)}$, 
$\gamma=-1$. 
\
\\

15) (0,0). $(F,F,F)$: $C_{(b_1|e_1)}$, $C_{(b_2|e_3)}$, $\gamma=-1$. 
$(FD,FD,F)$: $C_{(b_1|e_1)}$, $C_{(b_2|e_3)}$, $C_{(e_1|e_3)}$, 
$C_{(e_1|e_4)}$, $C_{(e_2|e_3)}$, $C_{(e_2|e_4)}$, $\gamma=-1$. 
$(FD,FD,FD)$: $C_{(b_1|e_1)}$, $C_{(b_2|e_3)}$, $C_{(e_2|e_5)}$, 
$C_{(e_4|e_5)}$, $\gamma=-1$.

\subsection*{\normalsize{\sl C.2. Reading the (super-)Higgs
mechanism directly from (\ref{sb1})--(\ref{slast})  }}

In Section 3 we saw how it is possible to interpret the GSO 
projections of the fermionic construction in terms of 
lattice shifts and twists and, in the light of the previous analysis, we are 
also able to understand them in terms of Higgs and super-Higgs mechanisms.
The Higgs mechanism is present whenever there are shifts due to modular 
coefficients $C_{(e_i|e_j)}=-1$: they translate in fact into $D$
projections.  There is a super-Higgs mechanism when there is a shift due
to modular coefficients $C_{(b_i|e_j)}=-1$ and/or $C_{(b_i|F)}=-1$
(by symmetric difference, the set $F$
can be seen to assign the boundary conditions 
of the sixth circle
of ${\cal T}^6$, which in the notation of Section 3, Eq. (\ref{zb}),
are $(\gamma,\delta)$). According to this 
interpretation, we see that the missing massless states still
belong to the string spectrum, and there are corners in moduli space in which
some or all of them become massless.
On each $N=4$ sector, besides the GSO projection that reduces the number
of states, starting from a maximum of sixteen, there is also in action
a GSO projection on the world-sheet chiralities, 
which determines whether such states are hyper- or vector multiplets.
By looking at formulae (\ref{sb1})--(\ref{slast}), which express the quantity
$N_V\pm N_H$ for each one of the 48 twisted (super)sectors, we can see that in
each sector the GSO projection splits into a product of three factors:
the first two factors, which determine whether a given twisted supersector
provides massless states or not, can be translated in terms of
lattice shifts. The modular coefficients entering the first factor
in square brackets determine the shift in the first circle of the 
corresponding untwisted torus,
while the shift in the second circle is determined by the coefficients
entering the second square brackets. The product of coefficients
after the square brackets translates into a shift in the twisted 
$T^4$.
A shift on a twisted lattice is directly related to the sign of $N_V-N_H$.  
The coefficients inside the square brackets 
therefore determine whether such states are massless or massive, while
the coefficients out
of the square brackets determine whether the massless states are hyper- or 
vector multiplets.

\noindent

\vskip 0.3cm
\setcounter{section}{0}
\setcounter{equation}{0}
\renewcommand{\theequation}{D.\arabic{equation}}
\section*{\normalsize{\centerline{\bf Appendix D: Classification
of partition functions}}}

The classification of the  ``partition functions'' 
can be easily carried out by observing that the situations
listed in (\ref{O}), (\ref{F}), (\ref{D}) and (\ref{DD}) 
are in a one-to-one correspondence with the number of massless
multiplets, no matter whether they are hypermultiplets or 
vector-multiplets, of the corresponding twisted sector.
The classification of ``partitions functions'' therefore amounts  
essentially to a complete account of these numbers.
There is, however, a subtlety, because this method
does not allow a distinction between (\ref{F}) and (\ref{FD}).
In fact, ``$D$'' can always be superposed to ``$F$'',
but not all the combinations are allowed, because 
the insertion of ``$D$'' makes sense only when it involves at least two 
circles belonging to two different tori.
The result is shown in Table D.1.
In this table we did not quote the constructions
that differ from the above by an exchange of $N_V$ and $N_H$ 
and/or by a permutation of the three planes.

From Table D.1, it easy to readread the number of supersymmetries
that are spontaneously broken. When 
the free action of a SUSY-breaking projection
appears in at most one plane (indicated by $F$ or $FD$), 
there is no spontaneous breaking of supersymmetry. 
When the free action involves two planes, there is a spontaneous 
breaking of $N=4$.
When finally the free action involves all the three planes,
there is spontaneous breaking of $N=8$.

\vspace{.7cm}
\[
\begin{tabular}{|| c || c | c | c || c | c | c || c | c | c ||}
\hline
$ (N_V,N_H)$ & $N_V^1$ & $N_V^2$ & $N_V^3$ & $N_H^1$ & $N_H^2$ & $N_H^3$ 
& plane 1 & plane 2 & plane 3 \rule[-.2cm]{0cm}{.8cm} \\ \hline
(48,0) & 16 & 16 & 16 & 0 & 0 & 0 & $O$ & $O$ & $O$ \rule[-.2cm]{0cm}{.7cm}\\ \hline
(28,4) & 8 & 8 & 12 & 0 & 0 & 4 & $D$ & $D$ & $O$ \rule[-.2cm]{0cm}{.7cm}\\ \hline
(16,16) & 8 & 8 & 0 & 8 & 8 & 0 & $O$ & $O$ & $F$ \rule[-.2cm]{0cm}{.7cm}\\ \hline
(24,0) & 8 & 8 & 8 & 0 & 0 & 0 & $D$ & $D$ & $D$ \rule[-.2cm]{0cm}{.7cm}\\ \hline
(18,6) & 4 & 4 & 10 & 0 & 0 & 6 & $DD$ & $DD$ & $O$ \rule[-.2cm]{0cm}{.7cm}\\ \hline
(12,12) & 4 & 8 & 0 & 4 & 8 & 0 & $D$ & $O$ & $FD$ \rule[-.2cm]{0cm}{.7cm}\\ \cline{2-10}
        & 4 & 4 & 4 & 4 & 4 & 4 & $D$ & $D$ & $D$ \rule[-.2cm]{0cm}{.7cm}\\ \hline
(16,4)  & 4 & 6 & 6 & 0 & 2 & 2 & $DD$ & $D$ & $D$ \rule[-.2cm]{0cm}{.8cm}\\ \hline
(14,2)  & 4 & 4 & 6 & 0 & 0 & 2 & $DD$ & $DD$ & $D$ \rule[-.2cm]{0cm}{.7cm}\\ \hline
(8,8)   & 2 & 2 & 4 & 2 & 2 & 4 & $DD$ & $DD$ & $D$ \rule[-.2cm]{0cm}{.7cm}\\ \cline{2-10}   
        & 4 & 4 & 0 & 4 & 4 & 0 & $D$ & $D$ & $F$ \rule[-.2cm]{0cm}{.7cm}\\ \cline{8-10}   
        &  &   &   &   &   &   & $D$ & $D$ & $FD$ \rule[-.2cm]{0cm}{.7cm}\\ \cline{2-10}   
        & 8 & 0 & 0 & 8 & 0 & 0 & $O$ & $F$ & $F$ \rule[-.2cm]{0cm}{.7cm}\\ \cline{8-10}  
        &  &   &   &   &   &   & $O$ & $FD$ & $FD$ \rule[-.2cm]{0cm}{.7cm}\\ \hline   
(12,0) & 4 & 4 & 4 & 0 & 0 & 0 & $DD$ & $DD$ & $DD$ \rule[-.2cm]{0cm}{.7cm}\\ \hline   
(6,6) & 2 & 2 & 2 & 2 & 2 & 2 & $DD$ & $DD$ & $DD$ \rule[-.2cm]{0cm}{.7cm}\\ \cline{2-10}   
      & 2 & 4 & 0 & 2 & 4 & 0 & $DD$ & $D$ & $FD$ \rule[-.2cm]{0cm}{.7cm}\\ \hline   
(9,3) & 3 & 3 & 3 & 1 & 1 & 1 & $DD$ & $DD$ & $DD$ \rule[-.2cm]{0cm}{.7cm}\\ \hline   
(4,4) & 4 & 0 & 0 & 4 & 0 & 0 & $D$ & $FD$ & $F$ \rule[-.2cm]{0cm}{.7cm}\\ \cline{8-10}   
      &  &   &    &   &   &   & $D$ & $FD$ & $FD$ \rule[-.2cm]{0cm}{.7cm}\\ \cline{2-10}   
      & 2 & 2 & 0 & 2 & 2 & 0 & $DD$ & $DD$ & $F$ \rule[-.2cm]{0cm}{.7cm}\\ \cline{8-10}   
      &  &   &    &   &   &   & $DD$ & $DD$ & $FD$ \rule[-.2cm]{0cm}{.7cm}\\ \hline   
(2,2) & 2 & 0 & 0 & 2 & 0 & 0 & $DD$ & $FD$ & $FD$ \rule[-.2cm]{0cm}{.7cm}\\ \hline   
(0,0) & 0 & 0 & 0 & 0 & 0 & 0 & $F$ & $F$ & $F$ \rule[-.2cm]{0cm}{.7cm}\\ \cline{8-10}   
      &  &   &   &   &   &   & $FD$ & $FD$ & $F$ \rule[-.2cm]{0cm}{.7cm}\\  \cline{8-10}  
      &  &   &   &   &   &   & $FD$ & $FD$ & $FD$ \rule[-.2cm]{0cm}{.7cm}\\ 
\hline   
\end{tabular}
\]

Table D.1: The contribution to the massless spectrum
and lattice sums in the $N=4$ sectors of the models.

\noindent

\vskip 0.3cm
\setcounter{section}{0}
\setcounter{equation}{0}
\renewcommand{\theequation}{E.\arabic{equation}}
\section*{\normalsize{\centerline{\bf Appendix E: Lattice integrals 
and threshold corrections}}}

We give below our notation and conventions for the usual
(2,2) and (2,2)-shifted lattice sums used in the text.
The $Z_2$-shifted (2,2) lattice sums are
\be
\Gamma_{2,2}^w (T,U) \ar{h}{g}  =
\sum_{\{p_{\rm L},p_{\rm R}\} \in \Gamma_{2,2}+w{h\over 2}}
{\rm e}^{-\pi i g \ell \cdot w }
q^{{1\over 2}p_{\rm L}^2} \bar{q}^{{1\over 2}p_{\rm R}^2}~, 
\label{zs}
\ee
where the shifts $h$ and projections $g$ take the values 0 or
1. Here, $w$ denotes the shift vector with components
$\left(a_1,a_2,b^1,b^2\right)$ and
$\ell \equiv \left(m_1,m_2,n^1,n^2\right)$. We have also introduced
the inner product\footnote{For
$w_1=\left({\vec a}_1,{\vec b}_1\right)$ and
$w_2=\left({\vec a}_2,{\vec b}_2\right)$,
the inner product is defined as
$w_1\cdot w_2= {\vec a}_1 {\vec b}_2+{\vec a}_2 {\vec b}_1$.}
\be
\ell \cdot w = \vec m \vec b + \vec a \vec n~,~~~
w^2 = 2  \vec a \vec b ~ ,
\label{inpr}
\ee
so that $a_I$ generates a winding shift in the $I$ direction, whereas
$b^I$ shifts the $I$th momentum. The vector $\ell$ is associated to
the $\Gamma_{2,2}$ lattice and therefore the vector associated to
the shifted lattice will be
\be
p\equiv \ell+w{h\over 2}~ .
\label{b3}
\ee
With these conventions, left and right momenta read:
\bs
\be
p_{\rm L}^2=
{\left\vert
U \left( m_{1}+a_1 {h \over 2}\right)-
\left( m_{2}+a_2 {h \over 2}\right)+
T \left( n^{1}+b^1 {h \over 2}\right)+
T U \left( n^{2}+b^2 {h \over 2}\right)
\right\vert^2\over 2 T_2 U_2}~ ,
\ee
\be
p_{\rm L}^2 - p_{\rm R}^2= 2
\left( m_{I}+a_I  {h \over 2}\right)
\left( n^{I}+b^I  {h \over 2}\right)~ .
\ee
\es
It is easy to check the periodicity properties ($h,g$ integers)
\be
Z_{2,2}^{w} \ar{h}{g}=
Z_{2,2}^{w} \ar{h+2}{g}=
Z_{2,2}^{w} \ar{h}{g+2}=
Z_{2,2}^{w} \ar{-h}{-g}
\label{b9}
\ee
as well as the modular transformations that the expression
\be
Z_{2,2}^{w} \ar{h}{g}=
{\Gamma_{2,2}^{w} \ar{h}{g}\over \vert \eta \vert^4}
\ee
obeys:
\be
\tau\to\tau+1 : \; \; Z_{2,2}^{w} \ar{h}{g}
\to
{\rm e}^{\pi i{w^2\over 2}{h^2 \over 2}} Z_{2,2}^{w}
\ar{h}{h+g}
\label{b8a}
\ee
\be
\tau\to-{1\over \tau} :\; \; Z_{2,2}^{w} \ar{h}{g} \to
{\rm e}^{-\pi i {w^2\over 2}{hg}} Z_{2,2}^{w} \ar{g}{-h} \, .
\label{b8b b8}
\ee
\label{b8}
The relevant parameter for these transformations is
\be
\lambda \equiv {w^2\over 2}=\vec a \vec b \, .
\label{b12}
\ee
From expressions (\ref{zs})  we learn that the
integers $a_I$ and $b^I$ are defined modulo 2, in the sense that adding
2 to any one of them amounts at most to a change of sign in
$Z_{2,2}^{w}{1\atopwithdelims[]1}$. Such a
modification is necessarily compensated by an appropriate one in
the rest of the partition function, in order to ensure modular
invariance; we are thus left with the same model. On the other
hand, adding 2 to $a_I$ or $b^I$ translates into adding a multiple of 2
to $\l$. Therefore,
although $\l$ can be any integer, only $\l=0$ and $\l=1$ correspond to
truly different situations. Only when $\lambda=0$, is the (2,2) block 
modular invariant by itself, when the sum over $(h,g)$ is taken.

In the case where there are two independent shifts $Z_2$, as in the 
cases we indicate
by $RR$, modular invariance requires also orthogonality of the two
shift vectors: $w_1 \cdot w_2=0$. The lattice sum, which we denote by
$\Gamma_{2,2}^{w_1,w_2}\ar{h_1,~h_2}{g_1,~g_2}$, satisfies the following
equalities:
\be
\Gamma_{2,2}^{w_1,w_2}\ar{h,~0}{g,~0}=
\Gamma_{2,2}^{w_1}\ar{h}{g}~,~~
\Gamma_{2,2}^{w_1,w_2}\ar{0,~h}{0,~g}=
\Gamma_{2,2}^{w_2}\ar{h}{g}~,~~
\Gamma_{2,2}^{w_1,w_2}\ar{h,~h}{g,~g}=
\Gamma_{2,2}^{w_{12}}\ar{h}{g}~,
\ee
where $w_{12} \equiv w_1+w_2$ reflects the action of the diagonal $Z_2$.
We refer to Appendix C of \cite{6auth} for a detailed discussion of
target space duality.
One of the issues, valid when $\lambda=0$, 
our case of interest, is that
a change in the lattice vector $w$, which preserves modular invariance
(i.e. $w^2/2=0~\hbox{mod}2$), amounts to an $SL(2,Z)$ transformation
performed on $T$ and/or $U$, and vice versa. 
We can therefore fix the lattice shift vectors and then derive the general 
result by $SL(2,Z)$ transformations. If we choose $w_1=(0,0,1,0)$, 
$w_2=(0,0,0,1)$, the first $Z_2$ translates the momenta of the first circle
(insertion of $(-1)^{m_1}$), the second $Z_2$ translates the momenta of the
second (insertion of $(-1)^{m_2}$)
\footnote{Notice that these are the same translations 
as were introduced when projecting with $D$ (\ref{d}).}. 
In this case the lattice sum reads:
\begin{eqnarray}
\Gamma_{2,2}{h_1,h_2\atopwithdelims[]g_1,g_2}&=&
\sum_{\vec m \vec n \in \mathbb Z}
(-1)^{m_1  g_1+m_2  g_2}
\exp
\Bigg\{
2\pi i\bar\t\left(
m_{1}\left( n^{1}+{h_1 \over 2}\right)+m_{2}\left( n^{2}+{h_2 \over
2}\right)
\right)\cr &&-\,
{\pi \t_2 \over T_2 U_2}
\left|T \left( n^{1}+{h_1 \over 2}\right) + TU\left( n^{2}+{h_2 \over
2}\right)+Um_1-m_2\right|^2
\Bigg\}\, ,
\label{lat22ex}
\end{eqnarray}
By performing a Poisson resummation over the momenta $(m_1,m_2)$,
we can express the shifted lattice in the Lagrangian formulation.
When $w=(w_1,w_2)=((0,0,1,0),(0,0,0,1))$, we have 
\be
Z_{2,2}\ar{h_1,~h_2}{g_1,~g_2}={1 \over |\eta|^4}~ \sum_{(m_1,n_1)}
\sum_{(m_2,n_2)}~Z_{2,2}\ar{n_1(h_1),~n_2(h_2)}{m_1(g_1),~m_2(g_2)}~,
\ee
where
\be
Z_{2,2}\ar{n_1(h_1),~n_2(h_2)}{m_1(g_1),~m_2(g_2)}=
(\hbox{Im} \tau)^{-1} \sqrt{{\rm det} G_{ij}}
\exp \left[ - \pi T_{ij}~ {(m_i+n_i \tau)(m_j+n_j \bar{\tau}) \over \hbox{Im} 
\tau } \right]~.
\label{zm}
\ee
In the above expression, the tensor $T_{ij}$ is defined as
\be
T_{ij}=G_{ij}+B_{ij}~,
\ee
where
\be
B_{ij}= \left ( \begin{array}{cc} 0 & -{\rm Im}T \\ {\rm Im}T & 0 
        \end{array} \right)~,~~~~~~ 
G_{ij}= { {\rm Im} T \over {\rm Im} U}
        \left ( \begin{array}{cc} 1 & {\rm Re}U \\ {\rm Re}U & |U|^2
        \end{array} \right)~,
\label{cbg}
\ee
and 
\be
m_i \in \mathbb Z+{g_i \over 2}~,~~~~n_j \in \mathbb Z+{h_j \over 2}~.
\ee
Relations (\ref{cbg}) can be inverted giving $T$ and $U$ as functions of
$B_{ij}$, $G_{ij}$:
\be
T=-B_{12}+i \sqrt{det G_{ij}}~,~~~~~~
U={G_{12} \over G_{11}}+i{\sqrt{det G_{ij}} \over G_{11}}~.
\label{ctu}
\ee
In terms of the metric $G_{ij}$ and the antisymmetric tensor $B_{ij}$,
the argument in the exponential of (\ref{zm}) becomes
\be
- \pi G_{ij} ~ {(m_i+n_i \tau)(m_j+n_j \bar{\tau}) \over 
\hbox{Im} \tau }+2 \pi i B_{ij}\, m_in_j~.
\ee
By using the identity:
\be
\vartheta \ar{a}{b}\bar{\vartheta} \ar{a}{b}=
{1 \over \sqrt{ 2 \hbox{Im} \tau}} ~\sum_{(m,n)}
{\rm e}^{- \pi~ {|m+n \tau|^2 \over 2 \hbox{{\tiny Im}} \tau }}
{\rm e}^{i \pi(am+bn+mn)}~,
\ee
it is easy to prove that the equality (\ref{gs}):
\be
\Gamma_{2,2}^{w}\ar{h_1,\;h_2}{g_1,\;g_2}=\sum_{a_1,b_1,a_2,b_2}
{\rm e}^{i \pi (a_1g_1+b_1h_1+h_1g_1)} {\rm e}^{i \pi (a_2g_2+b_2h_2+h_2g_2)}
\left| \vartheta \ar{a_1}{b_1}\vartheta \ar{a_2}{b_2} \right|^2~,
\label{cgs}
\ee
holds for $w=(w_1,w_2)=((0,0,1,0),(0,0,0,1))$ at the particular value of moduli
\be
T_0=i~,~~~~~~U_0=i~.
\ee
Notice also that this is the self-dual point.
\begin{center}

$\star~\star~\star$

\end{center}
We recall here the integrals of shifted lattice sums.
If there is no shift, we have \cite{dkl}:
\be
\int_{\cal F}\left(\Gamma_{2,2}(T,U) -1\right) 
=-\log \left(
T_2\left|\eta(T) \right|^4
U_2\left|\eta(U) \right|^4
\right)
- \log {8 \pi  {\rm e}^{1-\gamma}\over \sqrt{27}}\, .
\label{dkl}
\ee
In the case where there is only one $Z_2$ shift, as in expressions
(\ref{F}), i.e. when $w \equiv (w_1,0)$,
we have:
\be
\int_{{\cal F}}\left(\sump\Gamma^{w}_{2,2}\ar{h}{g} (T,U) -1\right)=
-\log
\left(
T_2\left|\vartheta_i(T) \right|^4
U_2\left|\vartheta_j(U) \right|^4
\right)
- \log {\pi  {\rm e}^{1-\gamma}\over 6\sqrt{3}}\, ,
\label{b15}
\ee
where the relation between the shift vector $w_1=(\vec{a},\vec{b})$  and
the pairs  $(i,j)$ is given in Table E.1:
\begin{center}
\begin{tabular}{| c || c | c || c | c | }
\hline
{\rm Case}&$\vec{a}$ & $\vec{b}$   & $i$ & $j$         \\ \hline
I   &$(0,0) $  & $(1,0)$   &  4 & 2    \\ \hline
II  &$(0,0) $  & $(0,1)$   &  4 & 4  \\ \hline
III &$(0,0) $  & $(1,1)$   &  4 & 3   \\ \hline
IV  &$(1,0) $  & $(0,0)$   &  2 & 4 \\ \hline
V   &$(0,1) $  & $(0,0)$   &  2 &  2  \\ \hline
VI  &$(1,1) $  & $(0,0)$   &  2 & 3 \\ \hline
VII &$(1,0) $  & $(0,1)$   &  3 & 4 \\ \hline
VIII&$(0,1) $  & $(1,0)$   &  3 & 2 \\ \hline
IX  &$(1,-1)$  & $(1,1)$   &  3 & 3 \\ \hline
\end{tabular}
\end{center}
\centerline{Table E.1: The nine physically distinct models with
$\l=0$.}
 
For the other cases, given in Eqs. (\ref{FD})-(\ref{DD}),
the integral is
obtained by taking the proper combination of (\ref{dkl})
and (\ref{b15}).
We collect here the results, including also the infrared 
running.
Modulo an integration constant, the result is:
\ba
I(O)~ & = & 
3 \log M_S^2/\mu^2 -3 \log (| \eta(T)|^4| \eta(U)|^4 T_2 
U_2)~,  \label{bo} \\
&& \nonumber \\
I^w(F)~ & = & 
\log M_S^2/\mu^2 - \log (| \vartheta_i(T)|^4| \vartheta_j(U)|^4 
T_2 U_2)~, \label{bf} \\
&& \nonumber\\
I^w(FD)~ & = & 
\log M_S^2/\mu^2 -1/2 ~\log (| \vartheta_{i}(T)|^4| 
\vartheta_{j}(U)|^4 T_2 U_2)+\nonumber \\ 
&&-1/2 ~\log (| \vartheta_{k}(T)|^4| \vartheta_{\ell}(U)|^4 
T_2 U_2)~ \label{bfer} \\
&& \nonumber \\
I^w(D)~ & = &
2\log M_S^2/\mu^2 - 3/2 ~\log (| \eta(T)|^4| \eta(U)|^4 
T_2 U_2)+ \nonumber \\
&&  - 1/2 ~\log (| \vartheta_i(T)|^4| \vartheta_j(U)|^4 T_2 U_2)~, \label{br} 
\\
&& \nonumber \\
I^w(DD) & = &
3/2~\log M_S^2/\mu^2 
- 3/4 ~\log (| \eta(T)|^4| \eta(U)|^4 T_2 U_2) \nonumber \\ 
&& - 1/4 ~\sum_{a=1}^3 ~\log (| \vartheta_{i_a}(T)|^4| \vartheta_{j_a}(U)|^4 
T_2 U_2)~ \label{brr}. 
\ea
In expression (\ref{bfer}), the first term
is due to the integration of a lattice with only one $Z_2$ shift, $w=(w_1,0)$,
and the dependence of the pairs $(i,j)$ 
on $w_1$ is given in Table E.1. The second term is due to the integration of 
a lattice with a shift specified by $w'=w_1+w_2$, where $w_1$ is the same as 
in the first term and $w_2$ is a vector in an independent direction.
Modular invariance requires $w_1^2=(w_1+w_2)^2=0$.   
The pair $(k,\ell)$ can be anyone of Table E.1, with the only
constraint $(k,\ell) \neq (i,j)$.
\
\\

In the case of two $Z_2$ shifts, as in (\ref{brr}), 
there is a sum of three terms, $a=1,2,3$, which refer 
respectively to the shifts given by the vectors $(w_1,w_2,w_1+w_2)$. 
The requirement of modular invariance 
reduces the number of possibilities to the following six:  
\begin{center}
\begin{tabular}{|c||c|c|}
\hline
{\rm Case}       & $w_1$         & $w_2$         \\ \hline
(\romannumeral1) & $(0,0,1,0) $  & $(0,0,0,1) $  \\ \hline
(\romannumeral2) & $(1,0,0,0) $  & $(0,1,0,0) $  \\ \hline
(\romannumeral3) & $(1,0,0,1) $  & $(0,-1,1,0)$  \\ \hline
(\romannumeral4) & $(1,0,0,0) $  & $(0,0,0,1) $  \\ \hline
(\romannumeral5) & $(0,0,1,0) $  & $(0,1,0,0) $  \\ \hline
(\romannumeral6) & $(0,0,1,1) $  & $(1,-1,0,0)$  \\ \hline
 
\end{tabular}
\end{center}
\centerline{Table E.2: The six physically distinct models with
$w_i^{\vphantom{1}} \cdot w_j^{\vphantom{1}} =0 \; \forall i,j = 1,2$.}

\noindent

\vskip 0.3cm
\setcounter{section}{0}
\setcounter{equation}{0}
\renewcommand{\theequation}{F. \arabic{equation}}
\section*{\normalsize{\centerline{\bf Appendix F: 
The helicity supertraces in the type II asymmetric orbifolds}}}
\noindent

As we discussed in Section 4,
the helicity supertraces are defined in terms of the four-dimensional helicity
$\lambda$ as 
\be
B_{2n} \equiv {\rm Str}~\lambda^{2n}~.
\ee
In the framework of string theory,
the physical four-dimensional helicity is 
the sum of the contributions of the left and right movers:
$\lambda=\l_{\rm L}+\bar \l_{\rm R}$.
The supertraces are computed by 
acting on the helicity-generating partition function
$Z(v,\bar{v})$ with the differential operators
that represent $\l_{\rm L} \left( \bar \l_{\rm R} \right)$:
\be
\l_{\rm L}={1 \over 2 \pi i} \partial_{v}~,~~~~~~~
\bar \l_{\rm R} =-{1 \over 2 \pi i} \partial_{\bar{v}}~.
\ee
In the type II asymmetric orbifolds of Section 5.3, the contribution of the
right-moving fermions cannot be cast directly in the form
(\ref{frv}) (with $H^2=G^2=0$). In order to compute the helicity supertraces
we must start from the expression
\be
Z_{\rm R}^{\rm F} \ar{a,H^{\rm o}}{b,G^{\rm o}}\left( \bar{v} \right)=
{(-)^{\bar{a}+\bar{b}+\bar{a}\bar{b}}
\over \bar{\eta}^4} \, \bar{\vartheta} \ar{\bar{a}}{\bar{b}} 
\left( \bar{v} \right)
\bar{\vartheta} \ar{\bar{a}}{\bar{b}} 
\bar{\vartheta} \ar{\bar{a}+H^{\rm o}}{\bar{b}+G^{\rm o}}
\bar{\vartheta} \ar{\bar{a}-H^{\rm o}}{\bar{b}-G^{\rm o}}~.
\label{ZaH}
\ee
The helicity-generating partition function reads
\ba
Z(v,\bar{v}) & = &
{ \xi(v) \bar{\xi}(\bar{v}) \over \left\vert \eta \right\vert^4}~
{1 \over 4}
\sum_{H^{\rm F},G^{\rm F}} \sum_{H^{\rm o},G^{\rm o}}
{ \Gamma_{6,6} \ar{H^{\rm F},H^{\rm o}}{G^{\rm F},G^{\rm o}}
\over \left\vert \eta \right\vert^{12}}
\nn \\
&& \times ~ 
{1 \over 2} \sum_{\bar{a},\bar{b}}
(-)^{\bar{a}G^{\rm F}
+\bar{b}H^{\rm F}+H^{\rm F}G^{\rm F}}
Z_{\rm L}^{\rm F} \ar{H^{\rm o}}{G^{\rm o}}\left( v \right)
Z_{\rm R}^{\rm F} \ar{\bar{a},H^{\rm o}}{\bar{b},G^{\rm o}}
\left( \bar{v} \right)~,
\ea
where $Z_{\rm L}^{\rm F} \ar{H^{\rm o}}{G^{\rm o}}\left( v \right)$
is the same as expression (\ref{flv}), with $H^2=G^2=0$
and arguments $(H^{\rm o},G^{\rm o})$ instead of $(H^1,G^1)$.

\subsection*{\normalsize{\sl The helicity supertrace $B_2$}}

The only non-vanishing contribution to $B_2$ can originate from 
the sectors of the orbifold for which $(H^{\rm o},G^{\rm o}) \neq (0,0)$.
In these sectors, there is a constraint, 
coming from the twisted boson character,
which is non-vanishing only for $(H^{\rm F},G^{\rm F})=(0,0)$
or $(H^{\rm F},G^{\rm F})=(H^{\rm o},G^{\rm o})$.
When $(H^{\rm F},G^{\rm F})=(0,0)$, we get 1/2 of the
contribution of one $N=(2,2)$ sector of the corresponding type IIA
symmetric orbifolds, which is zero
because of the non-complete saturation of the fermion zero modes.
When $(H^{\rm F},G^{\rm F})=(H^{\rm o},G^{\rm o})$,
we get an identical contribution.
In order to see this, we redefine the arguments in (\ref{ZaH}) as:
\be
\bar{a}+H^{\rm o} \to A~,~~~~~~
\bar{b}+G^{\rm o} \to B~.
\label{aA}
\ee
After this substitution,
we use the Riemann identity and recast the right-moving
fermion contribution 
\be
{1 \over 2}~ 
\sum_{\bar{a},\bar{b}}~(-)^{\bar{a}+\bar{b}+\bar{a}\bar{b}
+\bar{a}G^{\rm o}+\bar{b}H^{\rm o}+H^{\rm o}G^{\rm o}}
\bar{\vartheta} \ar{\bar{a}}{\bar{b}}(\bar v)~
\bar{\vartheta} \ar{\bar a}{\bar b} \bar{\vartheta} 
\ar{\bar{a}+H^{\rm o}}{\bar{b}+G^{\rm o}} 
\bar{\vartheta} \ar{\bar{a}-H^{\rm o}}{\bar{b}-G^{\rm o}}
\ee
as
\be
-~\bar{\vartheta} \ar{1}{1} \left( {\bar{v} \over 2} \right)~
\bar{\vartheta} \ar{1}{1} \left( -{\bar{v} \over 2} \right)~
\bar{\vartheta} \ar{1+H^{\rm o}}{1+G^{\rm o}} 
\left( {\bar{v} \over 2} \right)~
\bar{\vartheta} \ar{1-H^{\rm o}}{1-G^{\rm o}} 
\left( {\bar{v} \over 2} \right)~.
\ee
Also the contribution of this term therefore vanishes, due to the
non-complete saturation of the fermion zero modes.

\subsection*{\normalsize{\sl The helicity supertrace $B_4$}}

In these orbifolds, all the $N=4$ sectors, namely:
\begin{enumerate}
\item
the $N=(4,0)$
sector with $(H^{\rm o},G^{\rm o})=(0,0)$,
\item
the $N=(2,2)$ sector with
$(H^{\rm o},G^{\rm o})\ne (0,0)$ and $(H^{\rm F}, G^{\rm F})=(0,0)$,
\item
the $N=(2,2)$ sector with $(H^{\rm o},G^{\rm o})\ne (0,0)$ and $(H^{\rm F},
G^{\rm F})=(H^{\rm o},G^{\rm o})$,
\end{enumerate}
contribute to $B_4$.
In the first sector the contribution
is given by $\langle  \l_{\rm L}^4 \rangle$. We obtain 
\be
\langle  \l_{\rm L}^4 \rangle=
{3 \over 16}~{1 \over \bar{\eta}^{12}}~
\sum_{\bar{a},\bar{b}}\, \sum_{(H^{\rm F},G^{\rm F})}\,
(-)^{\bar{a}+\bar{b}+\bar{a}\bar{b}+\bar{a}G^{\rm F}
+\bar{b}G^{\rm F}+H^{\rm F}G^{\rm F}}
\bar{\vartheta}^4 \ar{\bar{a}}{\bar{b}} 
{1 \over 2^{n_D}} \sum_{\vec{H}^D,\vec{G}^D}
\Gamma_{6,6} 
\ar{H^{\rm F},~\vec{H}^D}{G^{\rm F},~\vec{G}^D}~ 
\ee
($n_D$ indicates the number of $D$-projections.)
This expression is a series that starts with a square-root pole:
\be
a_{-1}\, q^{-{1 \over 2}}+a_0+\ldots
\ee
The term $a_0$ gives the massless contribution, which turns out to be
constant in the full space of $T^{\rm As}$, $U^{\rm As}$,
the moduli in the vector multiplets, 
but not in the space of the moduli belonging to hypermultiplets.
At a generic point in moduli space, we have $a_0=6$. 
When $(H^{\rm o},G^{\rm o}) \neq (0,0)$, $B_4$ amounts to
$6 \langle \l_{\rm L}^2  \l_{\rm R}^2 \rangle$.
In this case, the arguments $(H^{\rm F},G^{\rm F})$, as we saw,
are constrained. One therefore proceeds   
as for the computation of $B_2$, by splitting the sum over 
$(H^{\rm F},G^{\rm F})$ into the two terms $(H^{\rm F},G^{\rm F})=(0,0)$
and $(H^{\rm F},G^{\rm F})=(H^{\rm o},G^{\rm o})$. 
After the same substitution of variables as in (\ref{aA}), one obtains
that the contribution of each one of the two terms is equal to
the contribution of one complex plane of the
symmetric orbifold. In these sectors, the arguments
$(\vec{H}^D,\vec{G}^D)$ are constrained as well:
$(H^D,G^D)=(0,0)$ or $(H^D,G^D)=(H^{\rm o},G^{\rm o})$. 
One therefore gets the various expressions:
\ba
6 \langle \l^2 \bar \l^2 \rangle & = &
36 \Gamma_{2,2}^{(3)} \ar{0~|~0}{0~|~0}~~~~~~~~~~~~
~~~~~~~~~~~~~~~~~~~~~~~~~~~~~~~N_V=16~;\\
&& \nn \\
&=& 12 \sump \left({1 \over 2} \Gamma_{2,2}^{(3)} \ar{0~|~0}{0~|~0}+
{1 \over 2} \Gamma_{2,2}^{(3)} \ar{0~|~h}{0~|~g}  \right)~~~~~~~~~N_V=8~; \\
&& \nn \\
&=& 12 \sump \left({1 \over 4} \Gamma_{2,2}^{(3)} \ar{0~|~0,0}{0~|~0,0}+
{1 \over 4} \Gamma_{2,2}^{(3)} \ar{0~|~h,0}{0~|~g,0} \right. \nn \\
&& ~~ \left. +~ {1 \over 4} \Gamma_{2,2}^{(3)} \ar{0~|~0,h}{0~|~0,g}+
{1 \over 4} \Gamma_{2,2}^{(3)} \ar{0~|~h,h}{0~|~g,g}  \right)~~~~~~~~~
N_V=4~; \\
&=& 12 \sump
\Gamma_{2,2}^{(3)} \ar{0~|~h}{0~|~g}
~~~~~~~~~~~~~~~~~~~~~~~~~~~~~~~~~~~~N_V=0~.
\ea

\end{document}